\documentclass[preprint,12pt]{elsarticle}
\usepackage{amssymb}
\usepackage{amsmath}
\usepackage{subcaption}
\usepackage{multirow}
\usepackage[dvipsnames]{xcolor}
\usepackage{diagbox}
\newcommand{\de}{\partial}

\renewcommand{\Re}{{Re}}
\newcommand{\Pe}{{Pe}}
\newcommand{\Ch}{{Ch}}

\newcommand{\We}{{We}}
\newcommand{\Ex}{{E_x}}
\newcommand{\PI}{{Pi}}
\newcommand{\El}{\beta_s}
\usepackage[bordercolor=black,linecolor=black,backgroundcolor=white,textwidth=3.5cm]{todonotes}

\graphicspath{{./figures_pdf/}}

\usepackage{xcolor}

\journal{Journal of Computational Physics}
\begin{document}
\begin{frontmatter}

\title{Coalescence of surfactant-laden drops \\ by Phase Field Method}
\address[label1]{Institute of Fluid Mechanics and Heat Transfer, Technische Universit\"at Wien}
\address[label2]{Dipartimento Politecnico di Ingegneria e Architettura, Universit\`a di Udine}
\author[label1,label2]{Giovanni Soligo}
\author[label1,label2]{Alessio Roccon} 
\author[label1,label2]{Alfredo Soldati}  \ead{alfredo.soldati@tuwien.ac.at}

\begin{abstract}
In this work, we propose and test the validity of a modified Phase Field Method (PFM), which was specifically developed for large scale simulations of turbulent flows with large and deformable surfactant-laden droplets.
The time evolution of the phase field, $\phi$, and of the surfactant concentration field, $\psi$, are obtained from two Cahn-Hilliard-like equations together with a two-order-parameter Time-Dependent Ginzburg-Landau (TDGL) free energy functional. 
The modifications introduced circumvent existing limitations of current approaches based on PFM and improve the well-posedness of the model.
The effect of surfactant on surface tension is modeled via an Equation Of State (EOS), further improving the flexibility of the approach.
This method can efficiently handle topological changes, i.e. breakup and coalescence, and describe adsorption/desorption of surfactant.
The capabilities of the proposed approach are tested in this paper against previous experimental results on the effects of surfactant on the deformation of a single droplet and on the interactions between two droplets.
Finally, to appreciate the performances of the model on a large scale complex simulation, a qualitative analysis of the behavior of surfactant-laden droplets in a turbulent channel flow is presented and discussed.
\end{abstract}

\end{frontmatter}


\section{Introduction}
Surfactants (surface active agents) are compounds which can strongly affect the phenomena occurring in a multiphase flow.
Surfactant molecules, composed of an hydrophobic tail and of a polar head, preferentially line up at the interface decreasing the surface tension of pure (clean) fluids.

The action of surfactants modifies the dynamics of the interface with important consequences on the overall behavior of the multiphase flow. For instance, they can strongly influence the number of droplets or bubbles which form in a mixture, the behavior of interface waves, atomization, coalescence and breakup phenomena. All these phenomena have an enormous impact on the outcome of a number of industrial and environmental applications \cite{Dobbs1989,sjoblom2005emulsions,takagi2011surfactant}.
The effect of surfactants is not limited to alter the value of the surface tension.
They can also generate local streaming via the action of tangential stresses at the interface (so-called Marangoni stresses \cite{scriven1960marangoni}), which arise whenever gradients of surface tension (i.e. gradients of surfactant concentration) are generated along the interfaces.

The efficient and accurate computational modeling of interfacial flows in the presence of surfactants is a challenging task, since surfactants affect the flow introducing non-uniform capillary and tangential stresses. 
In turn, the flow field advects the surfactant influencing its distribution and thus making the problem coupled.
From a numerical point of view, a coupled system of equations must be solved on an ever moving and deforming interface, which may undergo topological changes (in the considered case breakup and coalescence).
This problem imposes further complexity to the already challenging problem of computing the flow of bubbles or drops. In the following, the literature on numerical methods used to describe bubble/droplet laden flows \cite{Elghobashi2018}, which is vast, and the one dealing with multiphase flows in presence of surfactant, which to the best of our knowledge is more limited, will be briefly reviewed. 
Numerical methods used to simulate interfacial flows can be roughly divided in two categories: interface tracking and interface capturing methods.
Interface tracking methods use a separate grid or mesh to track the interface.
The most popular are the Front-Tracking (FT) method \cite{bayareh2011,JesusRPVS_2015,LuMT_2017,MuradogluT_2014,zhang2006front}, the Boundary Integral Method (BIM) \cite{dai2008,LoewenbergH_1997,StoneL_1990} and the Immersed Boundary Method (IBM) \cite{lai2008immersed,lai2010numerical}. 
These methods, initially developed for insoluble surfactants, have been then extended to soluble surfactants \cite{MuradogluT_2008,zhang2006front}.
While these approaches offer a good accuracy, handling of topological changes requires complex algorithms, especially when dealing with coalescence or breakup in three dimensions.
Interface-capturing methods are based on the use of an indicator function to represent implicitly the interface on an Eulerian grid; this greatly simplifies the discretization and the handling of topological changes.
Among the interface capturing methods, we can find the more commonly used Volume-Of-Fluid (VOF) \cite{HIRT1981,SCARDOVELLI99} and Level-Set (LS) \cite{Osher1988,sethian2003level}, and the relatively newer Phase Field Method (PFM) \cite{Roccon2017,scarbolo2015,Scarbolo2016}.
In the frame of VOF, approaches initially developed for insoluble surfactants \cite{bazhlekov2006numerical,drumright2004effect,james2004surfactant,renardy2002new} have been then extended to soluble surfactants and 3D flows \cite{alke20093d}.
In the frame of LS method, a possible approach has been proposed by Xu and Zhao \cite{xu2003eulerian} and then further improved to consider flow and contact line dynamics  \cite{Xu2006,xu2014level,Xu2012}.
Recently, alternative approaches, which combine interface capturing/tracking methods or use different frameworks, have been developed.
LS and FT have been used together \cite{ceniceros2003effects} and techniques based on the so-called Arbitrary Lagrangian-Eulerian (ALE) were proposed \cite{uzgoren2008marker,yang2007arbitrary,young2009influence}.
Considering other frameworks, Smooth-Particle Hydrodynamics (SPH)  \cite{adami2010conservative} and color-gradient Lattice Boltzmann (LB) \cite{farhat2011,gunstensen1991lattice} approaches are  available in literature.

In this paper a modified Phase Field Method \cite{Anderson1998,Jacqmin1999} for the simulation of interfacial flows with soluble surfactants is proposed.
The method, based on an interface capturing technique, represents the interface and the surfactant concentration using two order parameters, the phase field, $\phi$, and the surfactant concentration, $\psi$. 
Their behavior is determined by two Cahn-Hilliard-like equations; the minimization of a two-order-parameter Ginzburg-Landau free energy functional \cite{Gu2014,Komura1997,Laradji1992,toth2015} governs their diffusive component.
The two order parameters, $\phi$ and $\psi$, are Eulerian variables and, thus, efficient and massively parallel numerical solvers can be used.
In addition, the interface capturing approach allows for the implicit description of topological changes and of surfactant adsorption (from the bulk to the interface) and desorption (from the interface to the bulk) phenomena.
Compared to previous formulations \cite{Komura1997,toth2015}, the free energy functional has been modified in order to maintain a uniform thickness of the interfacial layer, thus reducing the grid requirements and improving the well-posedness of the model \cite{Engblom2013,Yun2014}.
The phase field method is coupled with Direct Numerical Simulation (DNS) of the Navier-Stokes (NS) equations.
Surface tension forces are computed using a geometrical approach, together with an Equations Of State (EOS) to describe the surfactant action.
This modification further improves the flexibility of the method proposed.
The solution of the system of coupled equations is obtained via a highly-parallel solver based on a pseudo-spectral discretization \cite{KIM1987,Peyret2002}.

The paper is organized as follows: in the next section the governing equations are presented, in Section~\ref{sec: nummethod} the pseudo-spectral method adopted is described and in Section~\ref{sec: simsetup} the simulation setup is introduced.
The results of the numerical simulations are presented and discussed in Section~\ref{sec: results} and conclusions are drawn in Section~\ref{sec: conclusion}.

\section{Governing equations}
\label{sec: mathmodel}
The dynamics of a multiphase flow with surfactant is modeled coupling direct numerical simulations of the Navier-Stokes equations with a phase field method to compute the interface dynamics and the surfactant concentration.
The phase field method, which we previously used to study the dynamics of large and deformable droplets in turbulent flows \cite{Roccon2017,scarbolo2015}, is here used in a two-order-parameter formulation to describe interfacial flows with surfactants.
In the following, the governing equations of the two order parameters, phase field $\phi$ and surfactant concentration $\psi$, will be derived and then coupled with continuity and Navier-Stokes (NS) equations to describe the hydrodynamics of the system.

\subsection{Modeling the interface and the action of surfactant}
We consider a ternary system composed of a soluble surfactant and two immiscible phases.
In the frame of the phase field method, the system is described using two order parameters.
The first order parameter, $\phi$, (phase field) is used to describe the interface.
The phase field is uniform in the bulk of the two phases ($\phi=\pm1$) and changes smoothly across the interface.
The second order parameter, $\psi$, is used to describe the surfactant concentration, which is uniform in the bulk of the phases and reaches a maximum at the interface, where surfactant molecules preferentially accumulate. 
The phase field and the surfactant concentration are governed by two Cahn-Hilliard-like equations (reported here and in the following in a dimensionless form, see \ref{sec: nondim} for further details):
\begin{equation}
\frac{\de \phi}{\de t}+\mathbf{u}\cdot \nabla \phi=\frac{1}{\Pe_\phi}\nabla \cdot (\mathcal{M}_\phi(\phi) \nabla \mu_\phi)\, ;
\label{phgen}
\end{equation}
\begin{equation}
\frac{\de \psi}{\de t}+\mathbf{u}\cdot \nabla \psi=\frac{1}{\Pe_\psi}\nabla \cdot (\mathcal{M}_\psi(\psi) \nabla \mu_\psi)\, .
\label{psgen}
\end{equation}
where ${\bf{u}}=(u,v,w)$ is the velocity vector, $\mu_\phi$ and $\mu_\psi$ are the two chemical potentials, $\mathcal{M}_\phi$  and $\mathcal{M}_\psi$ are the two mobilities (or Onsager coefficients) and $\Pe_\phi$ and $\Pe_\psi$ are the two P\'eclet numbers. 
The latter ones represent the ratio between convective and diffusive phenomena for the two order parameters.

The expression of the chemical potentials $\mu_\phi$ and $\mu_\psi$ is derived from a two-order-parameter Ginzburg-Landau free energy functional $\mathcal{F}[\phi, \nabla \phi, \psi, \nabla \psi]$.
The functional is modeled as the sum of five different contributions:
\begin{equation}
\mathcal{F}[\phi, \nabla \phi, \psi, \nabla \psi]=\int_\Omega ( f_0 +f_{mix} + f_\psi + f_1 + f_{E_x}) d \Omega \, ,
\end{equation}
where $\Omega$ is the domain considered. 
The first term, $f_0$, is the ideal part of the free energy and accounts for the tendency of the system to separate into two pure fluids; this phobic behavior is described by a double-well potential:
\begin{equation}
f_0=\frac{1}{4}(\phi-1)^2(\phi+1)^2 \, .
\end{equation}
The term $f_0$ exhibits two minima corresponding to the two stable fluid phases, $\phi=\pm 1$.
The second term, $f_{mix}$, is a non-local term (mixing energy) accounting for the energy stored in the interfacial layer (surface tension) defined as:
\begin{equation}
f_{mix}=\frac{\Ch^2}{2}|\nabla\phi|^2 \, .
\end{equation}
The Cahn number, $Ch$, determines the interfacial layer thickness.
The contributions $f_0$ and $f_{mix}$ are function only of the phase field $\phi$; their mathematical expressions match those used when a clean system (absence of surfactant) is considered \cite{Roccon2017,Scarbolo2013b,Scarbolo2013}.

The presence of surfactant is modeled with three contributions to the energy functional: an entropy term, $f_\psi$, an adsorption term, $f_1$, and a bulk term, $f_{\Ex}$.
The term $f_\psi$ expresses the entropy decrease obtained when surfactant is uniformly distributed in all the domain and it is defined as:
\begin{equation}
f_\psi=\PI \left[\psi\log\psi +(1-\psi)\log(1-\psi) \right] \, .
\end{equation}
This contribution restricts the value assumed by $\psi$ to the range between $\psi=0$ (no surfactant) and $\psi=1$ (saturation of surfactant).
The temperature-dependent parameter $\PI$ determines the surfactant diffusivity.
Increasing $\PI$, diffusion increases and a uniform surfactant concentration in all the domain is favored.
The term $f_1$ favors the adsorption of the surfactant at the interface; thanks to their amphiphilic character (hydrophilic head and hydrophobic tail), surfactant molecules preferentially gather at the interface exposing their heads towards the water phase and their tails towards the other phase. 
The original term ($\propto |\nabla \phi|^2$ \cite{Laradji1992,vanderSman2006}) has been modified and replaced by a polynomial expression \cite{Engblom2013}:
\begin{equation}
f_1=-\frac{1}{2}\psi(1-\phi^2)^2  \, .
\end{equation}
This choice widens the range of parameters in which the problem is well-posed. The last contribution, $f_{E_x}$, penalizes the presence of surfactant in the bulk of the two phases and is defined as:
\begin{equation}
f_\Ex=\frac{1}{2E_x}\phi^2\psi \, .
\end{equation}
This term has a relevant contribution in the bulk of the two phases ($\phi=\pm 1$); by opposite, it vanishes at the interface ($\phi\simeq 0$).
The parameter $E_x$ sets the bulk surfactant solubility.

\begin{figure}[t]
\centering
\includegraphics[width=0.75\textwidth]{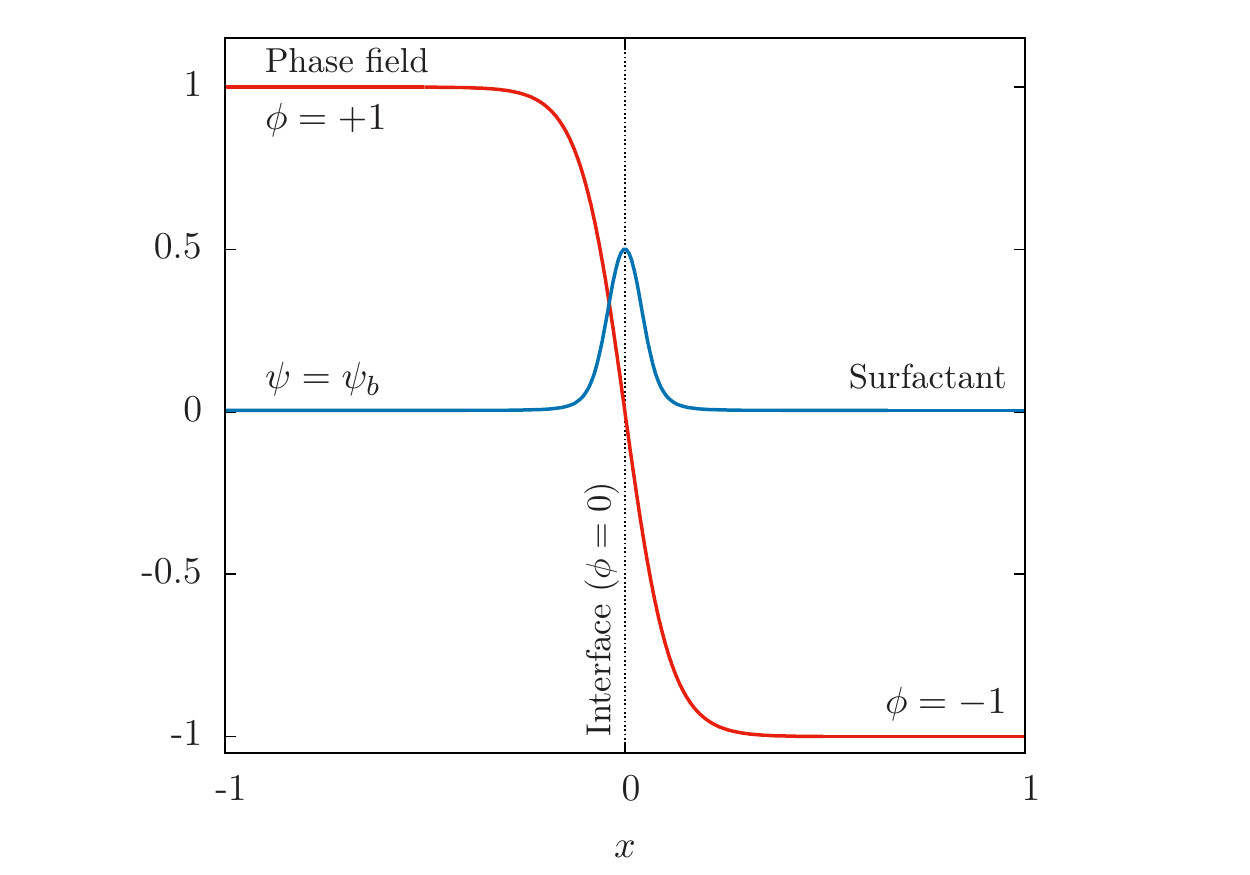}
\caption{
Equilibrium profile for the phase field $\phi$ (red line) and for the surfactant concentration $\psi$ (blue line). 
The phase field $\phi$ is uniform in the bulk of the two phases and it undergoes a smooth transition across the interface.
Likewise, the surfactant concentration $\psi$ is uniform in the bulk of the two phases, where $\psi=\psi_b$, and it increases at the interface, where surfactant molecules accumulate.
The surfactant concentration peak is found at $\phi=0$; its value depends on the parameters $\psi_b$, $Pi$ and $E_x$.}
\label{psiphieq}
\end{figure}

The expression of the chemical potentials is obtained taking the variational derivative of the free energy functional with respect to $\phi$ and $\psi$:
\begin{equation}
\mu_\phi=\frac{\delta \mathcal{F}}{\delta \phi}=\phi^3-\phi-\Ch^2\nabla^2\phi+ \overbrace{\Ch^2(\psi\nabla^2\phi+\nabla\psi\cdot\nabla\phi)+\frac{1}{\Ex}\phi\psi}^{C_{\phi \psi}} \, ;
\label{ciphipsi}
\end{equation} 
\begin{equation}
\mu_\psi=\frac{\delta \mathcal{F}}{\delta \psi}=\PI\log\left(\frac{\psi}{1-\psi}\right)-\frac{(1-\phi^2)^2}{2}+\frac{\phi^2}{2\Ex} \, .
\label{cipsipsi}
\end{equation}
The interfacial layer thickness, controlled by $\mu_\phi$, is influenced also by the surfactant concentration via the term referred as $C_{\phi \psi}$; this term can induce an unphysical behavior of the interface \cite{Yun2014}.
To restore the correct interfacial behavior, we neglect $C_{\phi \psi}$; in addition, surface tension forces are computed using a geometrical approach (which relies on the phase field $\phi$ to compute the interface curvature) together with an equation of state to describe the surfactant effect on surface tension \cite{Yun2014}.
This approach (compared to the thermodynamical one  \cite{Engblom2013,Laradji1992}) improves the flexibility of the method since surfactant action on surface tension is completely customizable.
Adopting these modifications, the equilibrium profiles of the two order parameters can be analytically derived. 
For the phase field $\phi$, the equilibrium profile is determined by the competition between  $f_0$ and $f_{mix}$.
At the equilibrium, $\mu_\phi=\mu_\phi^{eq}$ in the entire domain and from Eq.~(\ref{ciphipsi}) the following profile is obtained:
\begin{equation}
\phi(x)=\tanh \left( \frac{x}{\sqrt{2}Ch} \right) \, .
\label{eqprofphi}
\end{equation}
The phase field equilibrium profile reaches the values $\phi=\pm 1$ in the bulk of the phases ($x \rightarrow \pm \infty$) and undergoes a smooth transition following a hyperbolic tangent profile across the interface, as reported in Fig.~\ref{psiphieq}.

Likewise, the surfactant equilibrium profile can be deduced from Eq.~(\ref{cipsipsi}): at the equilibrium the surfactant chemical potential is constant throughout the entire domain. 
The surfactant equilibrium profile thus results in:
\begin{equation}
\psi(x)=\frac{\psi_b}{\psi_b+\psi_c(\phi) (1-\psi_b)} \, .
\label{eqprofpsi}
\end{equation}
The auxiliary variable $\psi_c$ is a function of the phase field solely:
\begin{equation}
\psi_c(\phi)=\exp\left[-\frac{1-\phi^2}{2\PI}\left(1-\phi^2+\frac{1}{\Ex}\right) \right]\, .
\end{equation}

At the equilibrium, the surfactant concentration is equal to $\psi=\psi_b$ in the bulk ($\phi=\pm1$) and has its maximum at the interface ($\phi=0$), Fig.~\ref{psiphieq}.
The maximum value of $\psi$ is influenced by the surfactant bulk concentration, $\psi_b$, and by the parameters $E_x$ and $Pi$.\\

The governing equations for the two order parameters can be completed defining the mobilities $\mathcal{M}_\phi$ and $\mathcal{M}_\psi$.
For the phase field, $\mathcal{M}_\phi$ is set constant \cite{Badalassi2003}, whereas for the surfactant is set to $\mathcal{M}_\psi(\psi)= \psi(1-\psi)$. The following two Cahn-Hilliard-like equations are obtained:
\begin{equation}
\frac{\de \phi}{\de t}+\mathbf{u}\cdot \nabla\phi=\frac{1}{\Pe_\phi}\nabla^2 (\phi^3-\phi-Ch^2 \nabla^2 \phi) \, ;
\label{eq:phi}
\end{equation}
\begin{equation}
\frac{\de \psi}{\de t}+\mathbf{u}\cdot \nabla\psi=\frac{\PI}{\Pe_\psi}\nabla^2 \psi + \frac{1}{Pe_\psi}\nabla\cdot\left[\mathcal{M}_\psi (\psi) \nabla\left(-\frac{(1-\phi^2)^2}{2}+\frac{\phi^2}{2\Ex}\right)\right] \, .
\label{eq:psi}
\end{equation}
These two equations describe the time evolution of the phase field $\phi$ and of the surfactant concentration $\psi$.

\subsection{Hydrodynamics}
\label{hydro}
The hydrodynamics behavior of the system is described coupling the two Cahn-Hilliard-like equations with continuity and Navier-Stokes (NS) equations.
This leads to a computational model able to accurately describe interfacial flows with surfactant.
In the most general case this approach can handle non-matched properties \cite{DING2007,Roccon2017}; density and viscosity are defined as a function of the phase field $\phi$.
In this work we want to focus on the effect of surfactant solely, so we considered two phases with matched density ($\rho=\rho_1=\rho_2$) and viscosity ($\eta=\eta_1=\eta_2$).
For the matched-property case, continuity and Navier-Stokes equations can be written as follows:
\begin{equation}
\nabla \cdot {\bf{u}}=0 \, ;
\label{co}
\end{equation}
\begin{equation}
\frac{\de\mathbf{u}}{\de t}+\mathbf{u}\cdot\nabla\mathbf{u}=-\nabla p +\frac{1}{\Re_\tau} \nabla^2 \mathbf{u}+\frac{3}{\sqrt{8}}\frac{\Ch}{\We}\nabla\cdot[\overline{\tau}_c f_\sigma(\psi)] \, ,
\label{nsfi}
\end{equation}
where ${\bf{u}}$ is the velocity field, $p$ is pressure and the last term of the right hand side is the interfacial term, which represents the surface tension forces \cite{popinet2018}.
These forces are calculated using a geometrical approach; in particular, the interface curvature is calculated from the phase field via the Korteweg stress tensor, $\overline{\tau}_c=|\nabla\phi|^2 \bf{I}-\nabla\phi\otimes \nabla \phi$ \cite{KORTEWEG1901}, while the surfactant action on surface tension is described with the equation of state $f_\sigma(\psi)$, described in Sec.~\ref{sec:eos}.
The interfacial term implicitly accounts for both the normal (capillary) and the tangential (Marangoni) components of surface tension forces and, indeed, it can be recasted as:
\begin{equation}
\nabla\cdot[\overline{\tau}_c f_\sigma(\psi)]= f_\sigma(\psi) \nabla \cdot \overline{\tau_c} + \nabla f_\sigma (\psi)  \cdot \overline{\tau_c}\, ,
\label{tanstress}
\end{equation}
where the terms on the right hand side are respectively the normal and tangential components of surface tension forces.
The latter one (tangential) vanishes when surface tension is uniform (surfactant is absent or uniformly distributed, $\nabla f_\sigma(\psi)=0$).
In the Navier-Stokes equations, two dimensionless groups are present: the shear Reynolds number, $\Re_\tau$, ratio between inertial and viscous forces and the Weber number, $\We$, ratio between inertial and surface tension forces.
In the definition of $\We$, the surface tension of a clean interface (referred in the following as $\sigma_0$) has been used as a reference value.

\subsection{Equation of state}
\label{sec:eos}
\begin{figure}[t]
\centering
\includegraphics[width=0.7\textwidth]{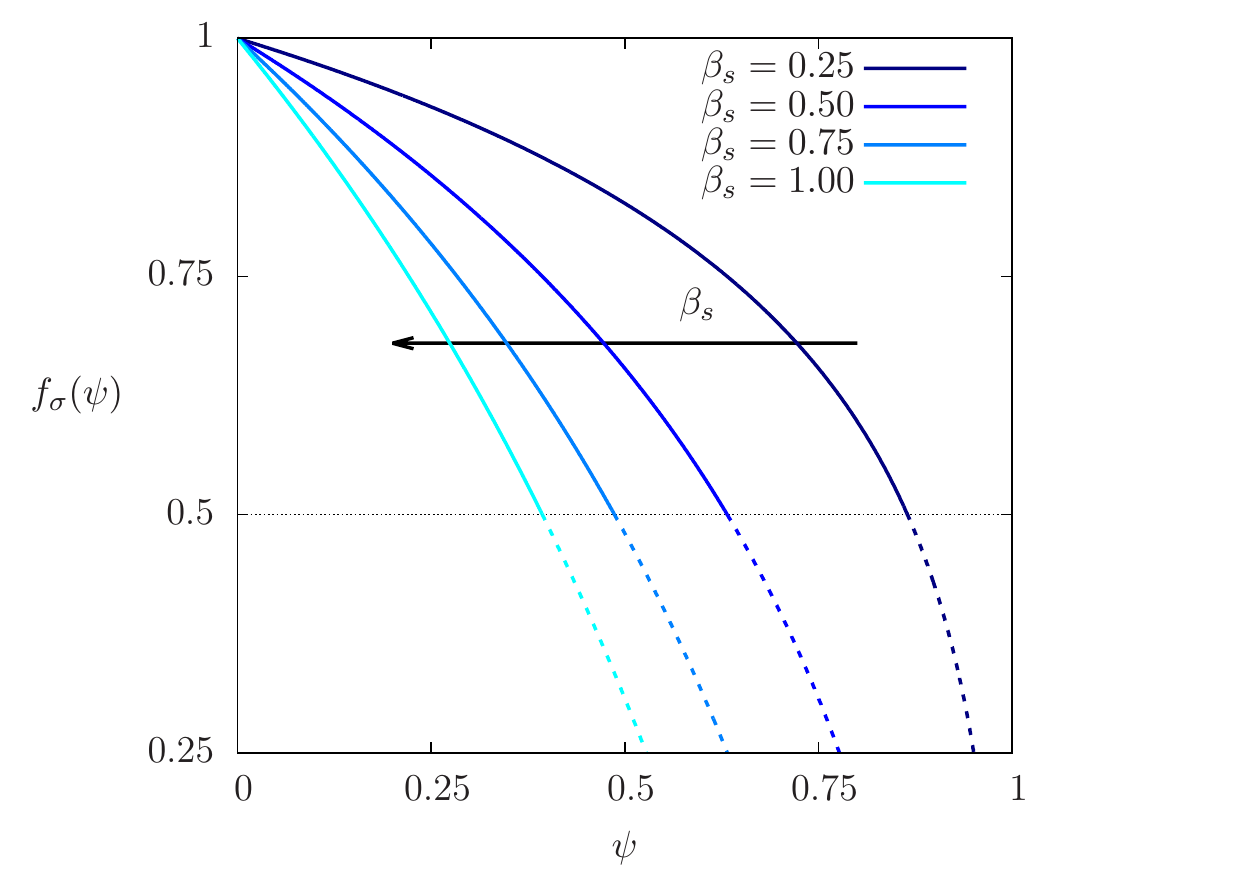}
\caption{Langmuir equation of state for different elasticity numbers $\El$, from $\beta_s=0.25$ (weak surfactant) to $\beta_s=1.00$ (strong surfactant). 
The Langmuir EOS gives an accurate description of the effect of surfactant down to $f_\sigma \simeq 0.5$; below this threshold, according to experimental observations \cite{chang1995adsorption}, surface tension does not decrease anymore (as the Langmuir EOS would predict) but it keeps constant.}
\label{eos}
\end{figure}

The surfactant action on surface tension is here described using an Equation Of State (EOS).
Experimental observations \cite{chang1995adsorption} show that, increasing the surfactant concentration, surface tension decreases until it approximately reaches half of its clean value, $\sigma(\psi)\simeq \sigma_0/2$; further increasing the surfactant concentration, surface tension keeps constant.
To describe this behavior, different EOSs have been proposed \cite{bazhlekov2006numerical,Pawar1996}; in this work we adopt a Langmuir EOS (Szyszkowski equation), valid in the limit of moderate surfactant concentrations.
The dimensionless Langmuir EOS is:
\begin{equation}
f_\sigma(\psi)=\frac{\sigma(\psi)}{\sigma_0}=1+\El\log\left(1-\psi\right) \, ,
\label{eq:sigma}
\end{equation}
where $\beta_s$ is the elasticity number, quantifying the strength of the surfactant.
In Fig.~\ref{eos}, the surface tension behavior predicted by the Langmuir EOS is shown for different elasticity numbers, $\beta_s$.
The equations of state predicts the correct surface tension decrease up to $f_\sigma=0.5$ (solid line); below this value the equation of state predicts a non-physical surface tension reduction (dashed line). In all the cases presented in this work, surface tension never reduced below the validity threshold of the surface tension EOS (the maximum local surfactant concentration was below $\sim0.4$).
For a fixed concentration, the higher is $\El$, the stronger is the surface tension reduction.

\section{Numerical Method}
\label{sec: nummethod}
The governing equations~(\ref{eq:phi})-(\ref{eq:psi})-(\ref{co}) and (\ref{nsfi}) are solved in a closed channel geometry using a pseudo-spectral method \cite{CanutoHQZ1988,HussainiZ1987,Peyret2002}. 
In particular, the equations are discretized using Fourier series in the streamwise and spanwise directions ($x$ and $y$) and Chebyshev polynomials along the wall-normal direction ($z$).
All the unknowns, velocity $\bf u$, phase field $\phi$ and surfactant concentration $\psi$, and their respective governing equations are Eulerian and have been solved on the same Cartesian grid; thus the coupling is straightforward and does not require any interpolation operation.
The governing equations have been recasted in a more compact form, collecting all the non-linear terms in the quantities $\bf S$, $S_\phi$ and $S_\psi$.
\begin{equation}
\nabla \cdot {\bf u}=0 
\end{equation}
\begin{equation}
\frac{\de \mathbf{u}}{\de t}=\mathbf{S}-\nabla p+\dfrac{1}{\Re_\tau}\nabla^2 \mathbf{u} 
\end{equation}
\begin{equation}
\frac{\de \phi}{\de t}=S_\phi+\dfrac{s}{\Pe_\phi}\nabla^2\phi -\frac{\Ch^2}{\Pe_\phi}\nabla^4\phi 
\label{ssphi}
\end{equation}
\begin{equation}
\frac{\de \psi}{\de t}=S_\psi+\dfrac{\PI}{\Pe_\psi}\nabla^2\psi 
\end{equation}

The terms $\bf{S}$, $S_\phi$ and $S_\psi$ are defined as follows:
\begin{equation}
\mathbf{S}=-\mathbf{u}\cdot\nabla\mathbf{u} +\dfrac{3}{\sqrt{8}}\dfrac{\Ch}{\We}\nabla\cdot\left[ \overline{\tau}_c f_\sigma (\psi) \right] \, ;
\end{equation}
\begin{equation}
S_\phi=-\mathbf{u}\cdot\nabla \phi+\dfrac{1}{\Pe_\phi}\left[\nabla^2\phi^3 -(1+s)\nabla^2\phi  \right]  \, ;
\label{sssphi}
\end{equation}
\begin{equation}
S_\psi=-\mathbf{u}\cdot\nabla\psi+\dfrac{1}{\Pe_\psi}\nabla\cdot\left[\psi(1-\psi)\nabla\left( -\dfrac{(1-\phi^2)^2}{2}+\dfrac{\phi^2}{2\Ex} \right)   \right] \, .
\end{equation}

The parameter $s$, equations (\ref{ssphi}) and (\ref{sssphi}), is a numerical coefficient used to perform the splitting of the Laplace operator. This technique improves the stability of the scheme \cite{Badalassi2003,YUE2004}; the coefficient $s$ is defined as:
\begin{equation}
s=\sqrt{\frac{4\Pe_\phi\Ch^2}{\Delta t}} \, .
\end{equation}

The governing equations are advanced in time using an IMplicit-EXplicit (IMEX) scheme. 
The linear diffusive term of the equations is integrated using an implicit scheme, whereas the non-linear term is integrated using an explicit scheme.
For the the Navier-Stokes equations, an Adams-Bashforth scheme is used for the non-linear terms while a Crank-Nicolson scheme is used for the linear term.
For the two Cahn-Hilliard-like equations, the non-linear terms are time-discretized using an Adams-Bashforth algorithm, while the linear terms are discretized using an implicit Euler algorithm.
The adoption of an implicit Euler method allows for the damping of unphysical high frequency oscillations that could arise from the steep gradients in the Cahn-Hilliard equations \cite{Badalassi2003,YUE2004}. 
At the first time step an explicit Euler method is used for the non-linear terms of all the equations.
At the generic time step $n$ the system of equations is discretized in time as follows:
\begin{equation}
\left\{
\begin{array}{l}
\dfrac{\de u^{n+1}}{\de x}+\dfrac{\de v^{n+1}}{\de y}+\dfrac{\de w^{n+1}}{\de z}=0 \\ [2ex]
\dfrac{\mathbf{u}^{n+1}-\mathbf{u}^n}{\Delta t}=\dfrac{3\mathbf{S}^{n+1}-\mathbf{S}^{n}}{2}-\nabla p+\dfrac{1}{\Re_\tau}\dfrac{\nabla^2 \mathbf{u}^{n+1} +\nabla^2 \mathbf{u}^{n}  }{2} \\ [2ex]
\dfrac{\phi^{n+1}-\phi^{n}}{\Delta t}=\dfrac{3S^{n+1}_\phi-S^{n+1}_\phi}{2}+\dfrac{s}{\Pe_\phi}\nabla^2\phi^{n+1} -\dfrac{\Ch^2}{\Pe_\phi}\nabla^4\phi^{n+1} \\ [2ex]
\dfrac{\psi^{n+1}-\phi^{n}}{\Delta t}=\dfrac{3S^{n+1}_\psi-S_\psi^{n}}{2}+\dfrac{\PI}{\Pe_\psi}\nabla^2\psi^{n+1} \\
\end{array}
\right.   \;.
\end{equation}

The solution of the Navier-Stokes equations with the pseudo-spectral method has been described previously \cite{CanutoHQZ1988}. 
However, since in the present case we solve a new set of equations, including several newly defined parameters, for the sake of clarity we report the details of the numerical scheme.
In this scheme, the Navier-Stokes equations are not directly solved but are rewritten in the so-called velocity-vorticity formulation. Instead of three $2^{nd}$ order equations for each component of the velocity, a $4^{th}$ order equation for the wall normal component of the velocity $w=\bf{u}\cdot \bf{k}$ (being $\bf{k}$ the versor of the wall normal direction) and a $2^{nd}$ order equation for the wall normal vorticity $\omega_z= (\nabla \times \bf{u}) \cdot \bf{k}$ are obtained.
In modal space, the full set of equations includes, in order: continuity, the definition of wall-normal vorticity, the wall-normal velocity transport, the wall-normal vorticity transport, the phase field transport and the surfactant concentration transport equations. 
A set of six independent equations for the six unknowns $\mathbf{u}=(u,v,w)$, $\omega_z$, $\phi$ and $\psi$ has to be solved. 
\begin{equation}
\left\{
\begin{array}{l}
\iota k_{x,i} u^{n+1}+ \iota k_{y,j} v^{n+1}+\dfrac{\textnormal{d} T_k}{\textnormal{d}z} w^{n+1}=0 \\ [2ex]
\omega_z^{n+1}=\iota k_{x,i} v^{n+1} -\iota k_{y,j} u^{n+1} \\ [2ex]
\begin{aligned}
\dfrac{\nabla^2 w^{n+1}-\nabla^2 w^n}{\Delta t}=&\dfrac{3}{2} (\nabla^2\mathbf{S}^n-\nabla(\nabla\cdot\mathbf{S}^n))\cdot\mathbf{k} - \\
& -\dfrac{1}{2}(\nabla^2\mathbf{S}^{n-1}-\nabla(\nabla\cdot\mathbf{S}^{n-1}))\cdot\mathbf{k}+\\
&+\dfrac{1}{2\Re_\tau}(\nabla^4\mathbf{u}^{n+1}+\nabla^4 \mathbf{u}^n)\cdot\mathbf{k}
\end{aligned}\\[2ex]
\dfrac{\omega_z^{n+1}-\omega_z^n}{\Delta t}=\dfrac{1}{2}\nabla\times \left(3\mathbf{S}^{n}-\mathbf{S}^{n-1}\right)\cdot\mathbf{k}+\dfrac{1}{2\Re_\tau}(\nabla^2\omega^{n+1} +\nabla^2 \omega^n)\cdot\mathbf{k} \\ [2ex]
\dfrac{\phi^{n+1}-\phi^n}{\Delta t}=\dfrac{1}{2}(3S_\phi^n-S_\phi^{n-1})+\dfrac{s}{\Pe_\phi}\nabla^2\phi^{n+1}- \dfrac{Ch^2}{\Pe_\phi}\nabla^4\phi^{n+1} \\ [2ex]
\dfrac{\psi^{n+1}-\psi^n}{\Delta t}=\dfrac{1}{2}(3S_\psi^n-S_\psi^{n-1})+\dfrac{\PI}{\Pe_\psi}\nabla^2\psi^{n+1} \\
\end{array}
\right.
\label{eq: discrsys}
\end{equation}
Superscripts denote the time step, being $n$ the current time step and $n+1$ the following one.
$k_{x,i}$ and $k_{y,j}$ are respectively the streamwise $i$-th and spanwise $j$-th wavenumbers; $\iota$ is the imaginary unit. $T_k$ is the $k$-th Chebyshev polynomial.
The equations are solved for each $(i,j,k)$ in $[1,N_x/2+1]\times[1,N_y]\times[1,N_z]$. 
The equations are solved separately: at first, the equations for the wall-normal velocity $w$ and vorticity $\omega_z$ are solved. 
Using the definition of vorticity and the continuity equation the new flow field ${\bf{u}}^{n+1}$ is obtained.
Then, the equations for the two order parameters $\phi$ and $\psi$ are solved.
The system of equations (\ref{eq: discrsys}) is the general formulation valid for a three dimensional case; when running 2D simulations the number of modes along one of the homogenous directions is limited to one (mean mode).

The numerical scheme presented above has been implemented in a Fortran 2003 proprietary code.
The code is parallelized using a 2D domain decomposition (pencil decomposition) strategy to divide the workload among the tasks, together with a pure-MPI paradigm to manage all the communications. 
Each task works on a fraction of the whole domain (pencil). When performing Fourier or Chebyshev transforms, all points in the transform direction are needed. Thus, to compute the transforms along the three different directions, the pencils have to be reorganized. This rearrangement is performed through MPI communications.

\section{Numerical simulations}
\label{sec: simsetup}

\subsection{Simulation Setup}

All the simulations aimed to analyze and benchmark the method have been performed on a 2D domain to examine in detail the role of the different parameters.
A final 3D, turbulent simulation, is also presented to highlight the capabilities of the method in dealing with large scale simulation of complex flows.
The 2D computational domain has dimensions $L_y \times L_z= 2 \pi \times 2$ and has been discretized using $N_y \times N_z= 512 \times 513$ collocation points along the streamwise and wall-normal direction.
The accurate description of the steep gradients at the interface requires a minimum of 5 grid points across the interface.
To meet this requirement the Cahn number, which determines the thickness of the thin interfacial layer, has been set to $Ch=0.02$.
Along the wall normal direction, where the grid is finer, up to 10 grid points are used.
The phase field P\'eclet number, $Pe_\phi$, has been set to $Pe_\phi=150$ following the scaling $Pe_\phi=3/Ch$ \cite{Magaletti2013,Yue2006} to achieve the sharp-interface limit.
For the surfactant, the P\'eclet number, which controls the diffusion, has been set to $\Pe_\psi=100$.
The parameters $\PI$ and $\Ex$ have been set respectively to $\PI=1.35$ and $\Ex=0.117$, the same values used by Engblom \textit{et al.} \cite{Engblom2013}; these two parameters influence the surfactant equilibrium profile, Eq.~(\ref{eqprofpsi}).
In all the simulations, $\PI$ and $\Ex$ were kept fixed, while the amount of surfactant was changed acting on the surfactant bulk concentration $\psi_b$.
We consider a shear flow configuration where the top and bottom walls move in opposite directions with velocity $v=\pm1$.
The shear Reynolds number is $\Re_\tau=0.1$ for the single droplet in shear flow and is increased to $\Re_\tau=0.5$ when the interaction between two droplets is considered. 
The initial flow field is a linear profile along the wall-normal direction for the streamwise component $v$; the other velocity components, $u$ and $w$, are set to zero. 
The phase field, $\phi$, and the surfactant concentration, $\psi$, are initialized with their equilibrium profile, equations~(\ref{eqprofphi}) and (\ref{eqprofpsi}).

\subsection{Boundary conditions}
A suitable set of boundary conditions has been imposed at the domain boundaries.
Specifically, at the walls, no-slip is enforced for the flow and a no-flux condition is used for both the phase field, $\phi$, and the surfactant concentration, $\psi$:
\begin{equation}
\left\{
\begin{array}{l}
\mathbf{u}(x,y,z=\pm1)=[0,\pm1,0] \\ [2ex]
\dfrac{\de \phi}{\de z} (x,y,z=\pm1)=0  \\ [3ex]
\dfrac{\de^3 \phi}{\de z^3} (x,y,z=\pm1)=0  \\ [3ex]
\dfrac{\de \psi}{\de z} (x,y,z=\pm1)=0  \\ [3ex]
\end{array}
\right .
\end{equation}

Along the streamwise and spanwise directions periodic boundary conditions are implicitly applied thanks to the Fourier discretization.
The boundary conditions imposed on $\phi$ and $\psi$ lead to a no-flux condition for the chemical potentials $\mu_\phi$ and $\mu_\psi$ and to the conservation of the two order parameters:
\begin{equation}
\frac{\partial}{\partial t} \int_\Omega \phi d \Omega =0 \quad , \quad
\frac{\partial}{\partial t} \int_\Omega \psi d \Omega =0 \, .
\label{mass}
\end{equation}
As a consequence, the total mass of the two phases and of the surfactant is conserved. 
Despite this, mass conservation of each of the two phases, $\phi=+1$ and $\phi=-1$, is not guaranteed \cite{Soligo2018,YUE2007} and some small mass leakages between these phases can be present (at most $\simeq$ 1\% in the simulations presented here). 

\section{Results}
\label{sec: results}
In this chapter, we will first benchmark our method performing the most commonly used tests, which include also a vis-a-vis comparison against previous experimental results \cite{dai2008,guido1998,hu2003,Hu2000,leal2004flow,pan2016,tretheway1999,wang2016}. 
Then we will examine on a qualitative basis the performances of the method on a fully turbulent, surfactant-laden multiphase flow.

\subsection{Single droplet in shear flow}
\label{single}

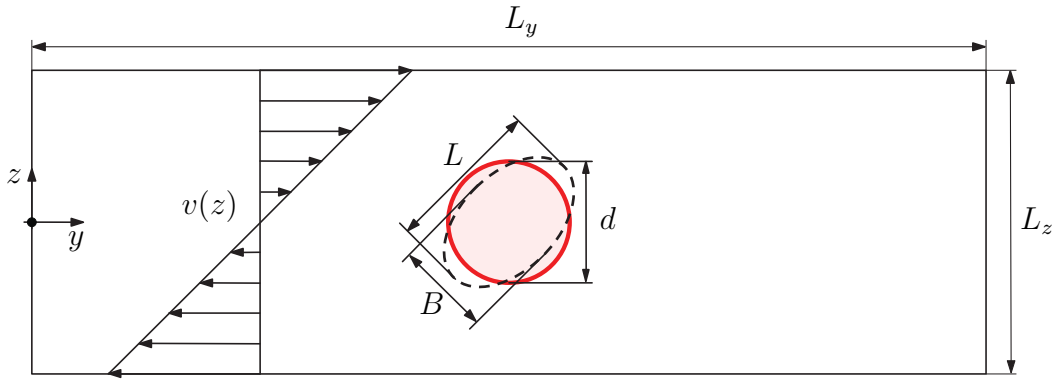
\begin{figure}[h!]
\center
\setlength{\unitlength}{0.0025\columnwidth}
\begin{picture}(400,140)
\put(6,0){\includegraphics[width=0.98\columnwidth, keepaspectratio]{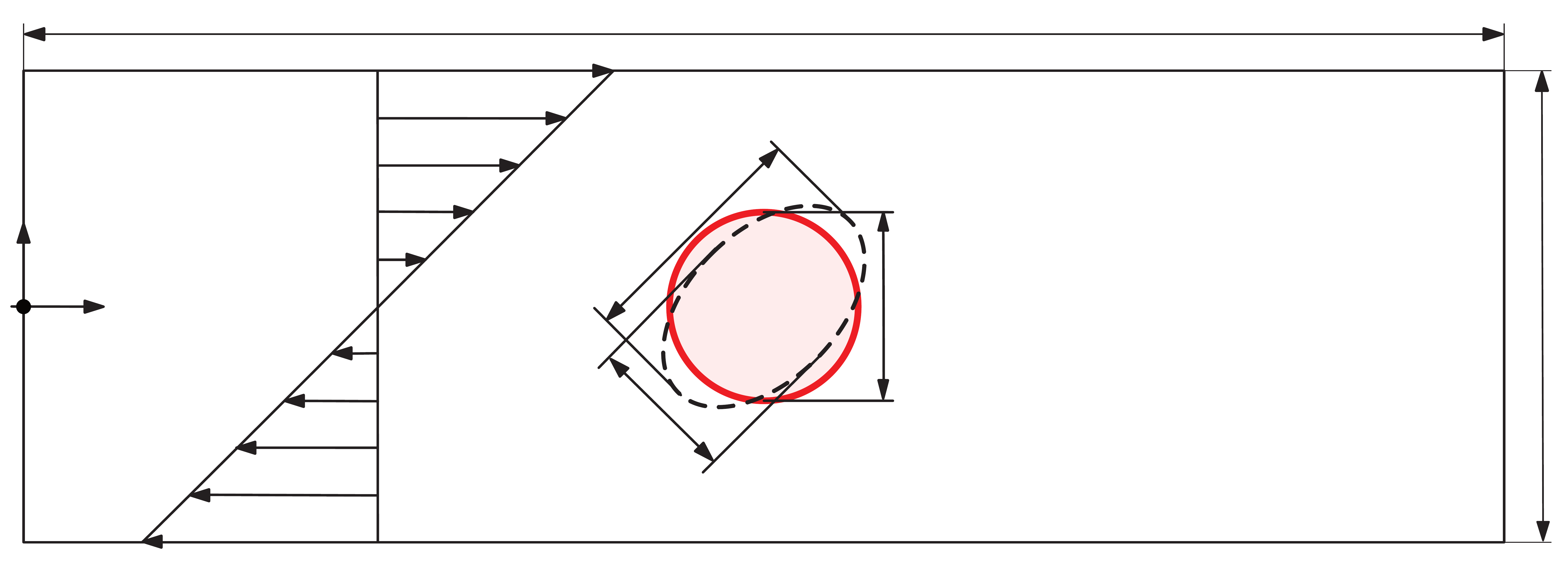}}
\put(2,80){$z$}
\put(26,56){$y$}
\put(171,87){$L$}
\put(162,29){$B$}
\put(195,140){$L_y$}
\put(395,62){$L_z$}
\put(232,62){\rotatebox{0}{$d$}}
\put(70,68) {$v(z)$}
\end{picture}
\caption{Sketch of the numerical setup used to analyze the deformation of a single droplet in shear flow. 
A circular droplet (red circle) is released in the center of the channel ($z_c=0$ and $y_c=\pi$); the channel has dimensions $L_y \times L_z= 2 \pi \times 2$. 
The shear flow  (linear velocity profile on the left) deforms the droplet until a new steady-state shape is obtained (dashed line). 
In this final configuration the droplet deformation parameter is computed.}
\label{fig: sh_flow}
\end{figure}

A circular droplet of diameter $d=0.8$ (red circle in Fig.~\ref{fig: sh_flow}) is released in the center of the channel ($z_c=0$ and $y_c=\pi$).
The shear flow deforms the droplet and advects surfactant along the interface.
After an initial transient the droplet reaches a new steady-state shape (dashed line in Fig.~\ref{fig: sh_flow}).
In this configuration, the major and minor axis $L$ and $B$ are measured and the deformation parameter $D$, defined as:
\begin{equation}
 D=\frac{(L-B)}{(L+B)}  \, ,
\label{deformation}
\end{equation}
is computed.
We start by considering the clean cases (absence of surfactant).
The final shape of the droplet is determined by the competition between viscous and surface tension forces; viscous forces try to elongate the droplet, while surface tension forces try to restore the circular shape.
The ratio between these forces is expressed by the capillary number $Ca$:
\begin{equation}
Ca=\frac{\We}{\Re_\tau}\frac{d}{2h} \, .
\label{capi}
\end{equation}
The capillary number is defined using the droplet radius as length scale, thus the rescaling factor $d/2h$, where $h$ is the channel half height.
We consider four different $Ca$, from $Ca=0.0625$ (highest surface tension) up to $Ca=0.250$ (lowest surface tension).

In order to compare the results with the analytic relation \cite{shapira1990,TAYLOR1934} only low $Ca$ have been considered.
Indeed, Taylor analytic relation considers 3D droplets which undergo limited deformations (low $Ca$); however, it was shown that this analytic formula well predicts also the deformation of 2D droplets for sufficiently low capillary numbers \cite{afkhami2009,yue2006phase}: at low $Ca$ the influence of the third dimension (normal to the velocity-velocity gradient plane) is negligible. As we restricted our simulations to low $Ca$ cases, we do expect a good agreement between the analytic relation and our results. 

\begin{figure}[t]
\center
\setlength{\unitlength}{0.0025\columnwidth}
\begin{picture}(400,200)
\put(50,0){\includegraphics[width=0.75\columnwidth,keepaspectratio]{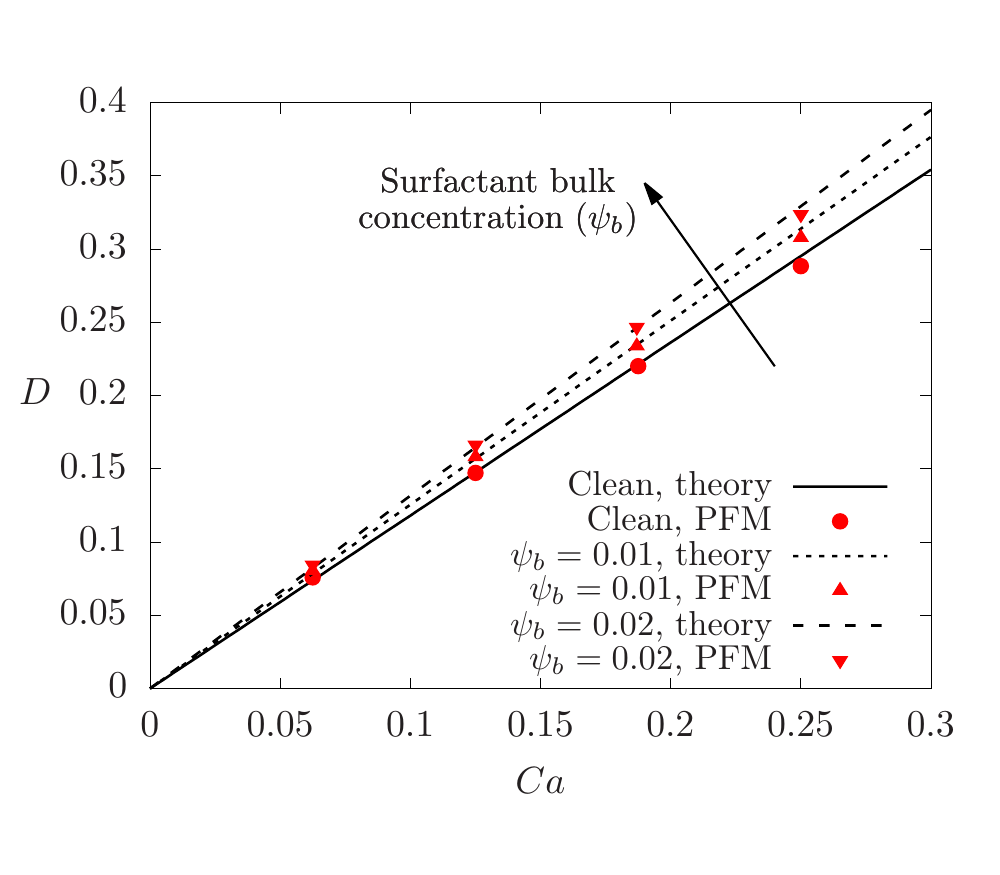}}
\end{picture}
\caption{
Comparison of numerical and analytic results for the deformation parameter $D$; analytic results for a clean droplet are reported with a solid black line, whereas numerical results are identified by red circles. 
For $\psi_b=0.01$, a double-dashed line identifies the analytic results and the numerical ones are identified by upward red triangles.
Similarly, for $\psi_b=0.02$, a dashed line identifies the theoretical results and the numerical ones are identified by downward red triangles.}
\label{dirty}
\end{figure}

In Fig.~\ref{dirty}, we compare our results with the ones predicted by the analytic relation developed by Taylor \cite{TAYLOR1934} and corrected by Shapira and Haber \cite{shapira1990} for confinement effects, which is the following:
\begin{equation}
D=\frac{35}{32}Ca\left[1+C_{SH}\frac{3.5}{2} \left(\frac{d}{4h}\right)^3\right] \, ,
\label{sheq}
\end{equation}
where $C_{SH}$ is a numerical coefficient equal to 5.6996 \cite{shapira1990}.
For all the cases considered, the numerical results (red circle markers) are in good agreement with the predictions of the analytic relation (black solid line).
At the highest $Ca$, the analytic relation slightly over-predicts the numerical result; this could be addressed to the limitation of the analytic relation (valid for low deformations). 

When surfactant is taken into account, the droplet shape is influenced by three new additional effects: $(i)$~surfactant decreases the average surface tension; $(ii)$~surfactant accumulates on the droplet tips producing non-uniform capillary forces; $(iii)$~inhomogeneous surfactant distribution gives rise to tangential stresses at the interface. 
The resulting outcome has been investigated considering, for each $Ca$ tested before, two further cases with surfactant bulk concentrations $\psi_b=0.01$ and $\psi_b=0.02$ and an elasticity number $\El=0.50$.
To compare the results obtained against the analytic relation \cite{shapira1990}, an effective capillary $Ca_e=(\sigma_0 /\sigma_{av}) Ca$ is used to compute the theoretical value of the deformation, where $\sigma_{av}$ is the average surface tension. 
The effective capillary accounts for the average surface tension reduction \cite{Stone1994,StoneL_1990}.

The results obtained from these new cases have been reported and compared against the analytic relation in Fig.~\ref{dirty}.
For both the surfactant bulk concentrations considered the analytic relation well predicts the results obtained from the simulations.
At the highest $Ca$, the analytic relation slightly over-predicts the numerical result, as previously noticed for the clean case.
Interestingly, the three surfactant-induced effects offset each other and, using $Ca_e$ in Eq.~(\ref{sheq}), a good prediction of the droplet deformation is found.
The numerical results obtained are in good agreement with the predictions of the analytic relation of Taylor \cite{TAYLOR1934} corrected by Shapira and Haber \cite{shapira1990}, also when the surfactant-laden cases are considered.
In addition, our results show a common trend with previous numerical  \cite{booty2010,farhat2011,frijters2012,khatri2011,teigen2011}  and experimental works  \cite{hu2003,Hu2000,tretheway1999}.
However, a direct comparison with these works is not possible since results are strongly affected by the surfactant type (soluble/insoluble), strength (equation of state and elasticity number) and loading (average concentration) employed.
\subsection{Droplet-droplet interaction in shear flow}
\label{double}
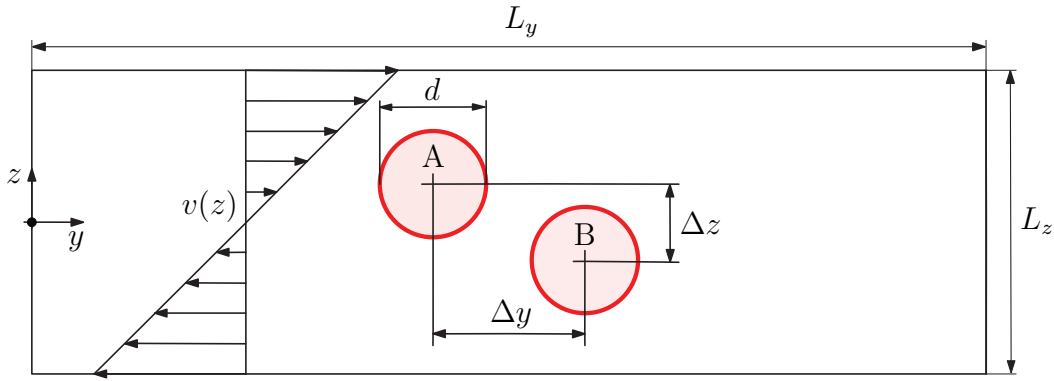
\begin{figure}
\center
\setlength{\unitlength}{0.0025\columnwidth}
\begin{picture}(400,140)
\put(6,0){\includegraphics[width=0.98\columnwidth, keepaspectratio]{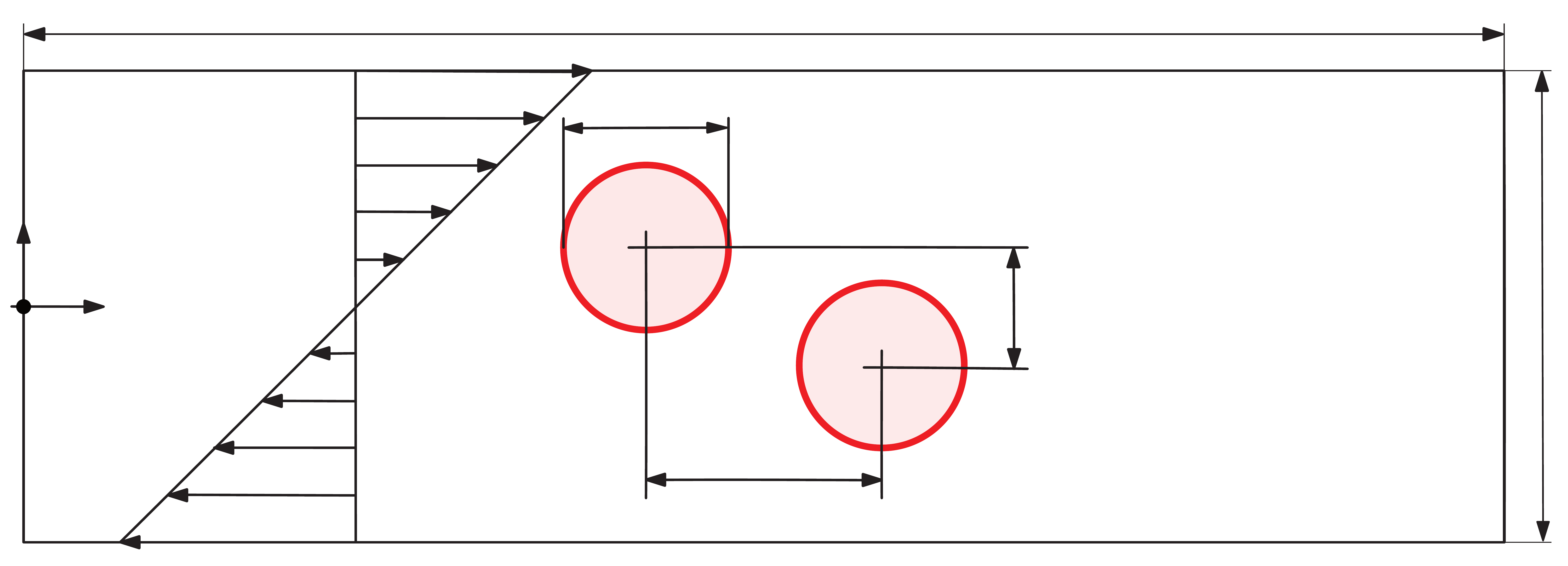}}
\put(2,80){$z$}
\put(26,56){$y$}
\put(164,112){\rotatebox{0}{$d$}}
\put(263,60){\rotatebox{0}{$\Delta z$}}
\put(189.5,26){\rotatebox{0}{$\Delta y$}}
\put(163,86){A}
\put(222,56){B}
\put(70,68) {$v(z)$}
\put(195,140){$L_y$}
\put(395,62){$L_z$}
\end{picture}
\caption{Sketch of the simulation setup used to analyze the effect of the surfactant on the droplet-droplet interaction. 
The droplets have a diameter $d=0.7$ and are separated by a distance $\Delta y=1$ and $\Delta z=0.5$.
The centers of the droplets are located at $y_c=\pi \mp 0.5$ and $z_c=\pm 0.25$; the channel size is $L_y \times L_z= 2 \pi \times 2$.} 
\label{problem}
\end{figure}
Surfactant effects on the interaction between two droplets have been analyzed considering two circular droplets of diameter $d=0.7$ in shear flow.
The droplets centers are located at $y_c=\pi \mp \Delta  y/2$ and $z_c=\pm \Delta z /2$, Fig.~\ref{problem}.
The shear flow drives the droplets towards each other: droplet A has a positive mean velocity (moves from left to right), while droplet B has a negative mean velocity (moves from right to left).
After the initial approaching stage, a thin liquid film forms between the droplets. 
If the liquid film thickness decreases below a critical threshold, the attractive van der Waals forces draw the interfaces even closer and the two droplets coalesce \cite{ChenCM_2009,dai2008,de2013effect,ha2003effect,leal2004flow,xu2011numerical,YUE2005}. 
The presence of a surfactant can drastically affect the interaction: indeed, the increased deformability and the tangential stresses at the interface hamper the draining of the thin liquid film and alter the interaction outcome \cite{dai2008}.
We analyze the effects of the surfactant considering different surfactant bulk concentrations, from $\psi_b=0.1 \times 10^{-2} $ (lowest concentration) to $\psi_b=1.0 \times 10^{-2}$ (highest concentration) and elasticity numbers, from $\El=0.125$ (weak effect on surface tension) up to $\El=1.00$ (strong effect on surface tension).

\subsubsection{Outcome of the interaction}
\begin{figure}[t]
\center
\setlength{\unitlength}{0.0025\columnwidth}
\begin{picture}(400,220)
\put(0,-20){\includegraphics[width=1.00\columnwidth, keepaspectratio]{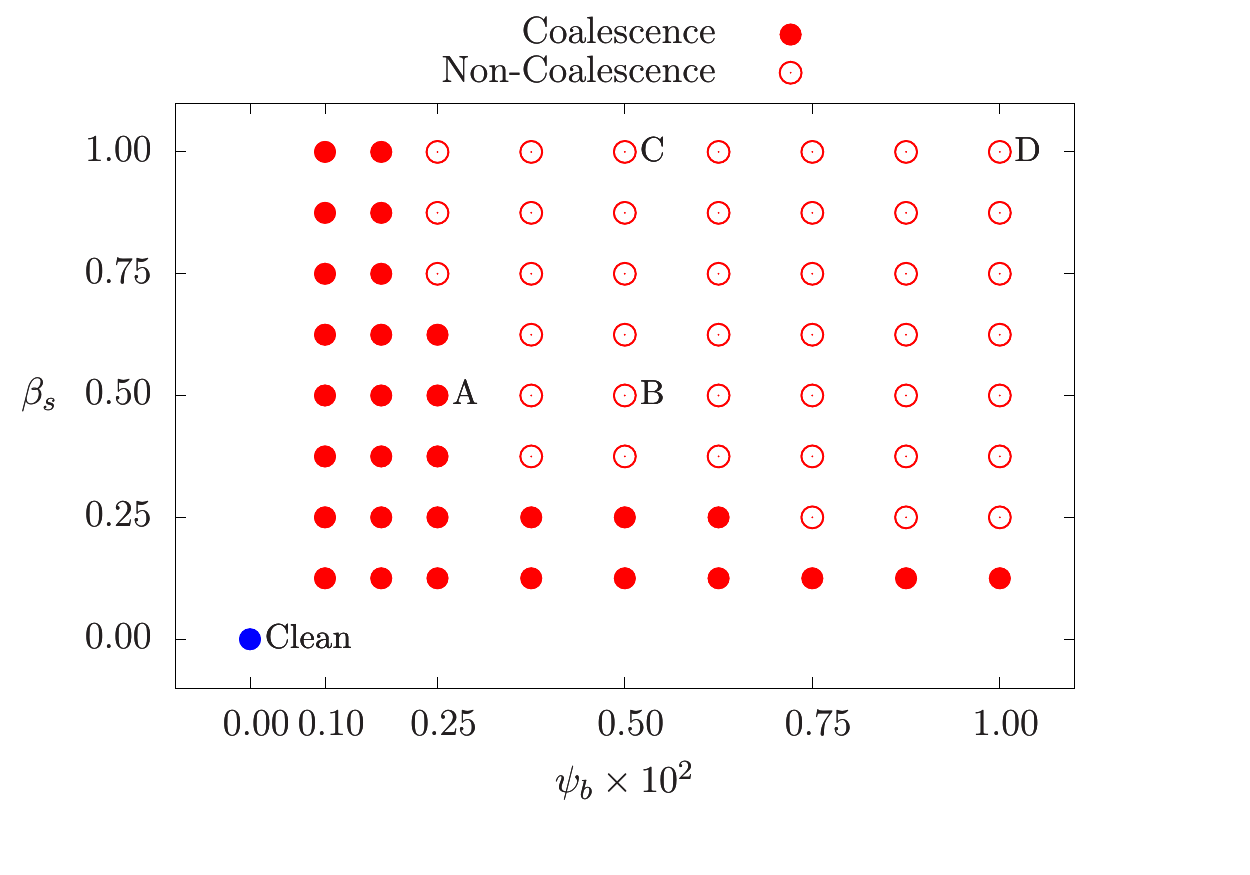}}
\end{picture}
\caption{Outcome of the droplet-droplet interaction for different surfactant bulk concentrations $\psi_b$ and elasticity numbers $\beta_s$.
A filled dot identifies a coalescence while an empty dot identifies a non-coalescence. 
The blue dot refers to the clean case (absence of surfactant).
Coalescence can be prevented by increasing the surfactant bulk concentration $\psi_b$ or the elasticity number $\beta_s$.
Simulations used as a reference in the following have been labelled (Clean, A, B, C, D).}
\label{map}
\end{figure}
\begin{figure}
\center
\setlength{\unitlength}{0.0025\columnwidth}
\begin{picture}(400,460)
\put(0,0){\includegraphics[width=1.00\columnwidth, keepaspectratio]{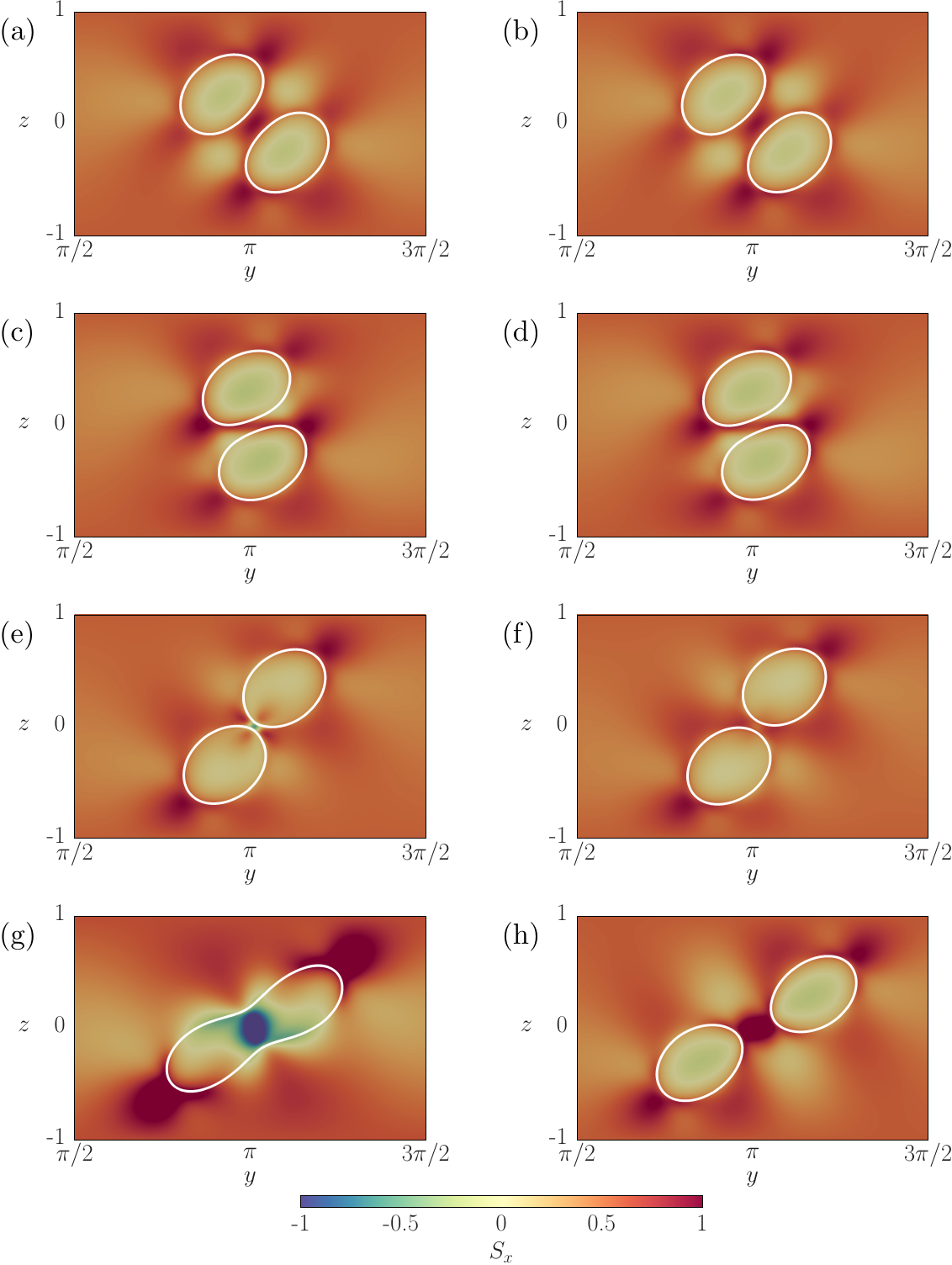}}
\end{picture}
\caption{Time evolution of the two droplets during the interaction. 
The left column, panels (a)-(c)-(e)-(g), refers to simulation A ($\beta_s=0.5$ and $\psi_b=0.25 \times 10^{-2}$) while the right column, panels (b)-(d)-(f)-(h), refers to simulation C ($\beta_s=1.0$ and $\psi_b=0.5 \times 10^{-2}$).
The white solid line shows the instantaneous droplet interface, iso-contour $\phi=0$.
Droplets coalescence occurs for simulation A; by opposite surfactant prevents the coalescence for case C. 
On the background, the strain rate $S_x= (\partial v/ \partial z + \partial w/\partial y)/2$ is plotted.}
\label{tphi}
\end{figure}
\begin{figure}
\center
\setlength{\unitlength}{0.0025\columnwidth}
\begin{picture}(400,460)
\put(0,0){\includegraphics[width=1.00\columnwidth, keepaspectratio]{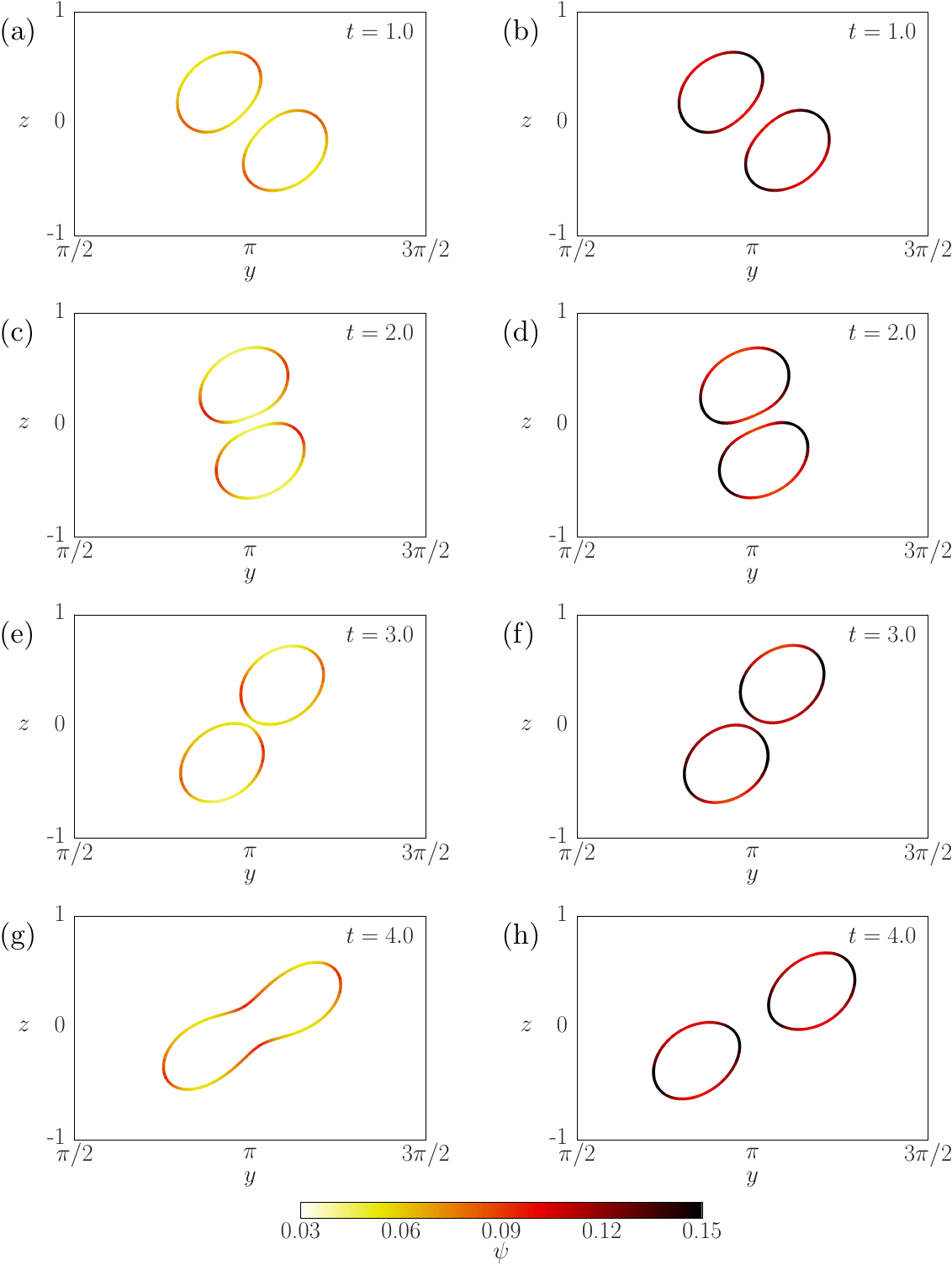}}
\end{picture}
\caption{Time evolution of the surfactant concentration at the droplet interface, iso-contour $\phi=0$, during the droplet-droplet interaction.
The left column, panels (a)-(c)-(e)-(g), refers to simulation A ($\beta_s=0.5$ and $\psi_b=0.25 \times 10^{-2}$) while the right column, panels (b)-(d)-(f)-(h), refers to simulation C ($\beta_s=1.0$ and $\psi_b=0.5 \times 10^{-2}$).
For case A, surfactant concentration is lower (lower $\psi_b$); by opposite for case C, surfactant concentration is higher (higher $\psi_b$).
For both cases, surfactant accumulates at the droplets tips.
For case A, after the coalescence, surface tension forces reshape the droplet, and surfactant redistributes.}
\label{tpsi}
\end{figure}
The most striking surfactant effect can be appreciated from the outcome of the droplet-droplet interaction: a map of the interaction outcomes for different combinations of the surfactant bulk concentration and of the elasticity number is reported in Fig.~\ref{map}. 
A filled dot identifies a coalescence event, while an empty dot marks a non-coalescence.
For the clean case ($\psi_b=0$ and $\beta_s=0$, blue dot), the two droplets coalesce: the absence of tangential stresses at the interface and the low deformability allow for the draining of the thin liquid film and the subsequent merging.
When surfactant is added, at the two lowest surfactant bulk concentrations coalescence occurs for each elasticity number considered: surfactant concentration is too low to have a noticeable effect.
Increasing the surfactant bulk concentration, from $\psi_b=0.37 \times 10^{-2}$ up to $\psi_b=0.62 \times 10^{-2}$, surfactant alters the interaction outcome.
For any elasticity number greater than $\beta_s=0.25$ coalescence is prevented, as for instance for the cases labelled as B and C.
The minimum elasticity number that prevents coalescence further reduces increasing the surfactant bulk concentrations: in the range $\psi_b\in [0.75\times10^{-2} , 1.00\times10^{-2}]$, coalescence occurs only for $\beta_s=0.125$.
Overall, an increase of either the surfactant bulk concentration, either the elasticity number prevents coalescence.

To give a better insight of the droplet-droplet interaction, the time evolution of the system for the cases labelled in Fig.~\ref{map} as A ($\beta_s=0.5$, $\psi_b=0.25 \times 10^{-2}$) and C ($\beta_s=1.0$, $\psi_b=0.5 \times 10^{-2}$) is reported in Figs.~\ref{tphi}-\ref{tpsi}.
In Fig.~\ref{tphi}, to highlight the different stages of the interaction, the interface of the drops and the strain rate $S_x= (\partial v/ \partial z + \partial w/\partial y)/2$ have been reported. 
Likewise, in Fig.~\ref{tpsi}, the surfactant concentration at the interface ($\phi=0$) has been reported; surfactant concentration in the bulk is not represented since it has an uniform value equal to $\psi_b$ (see Fig.~\ref{psiphieq}).
In both figures, the left column refers to case A while the right column to case C; time increases from the top to the bottom.

At time, $t=1.0$, Fig.~\ref{tphi}(a)-(b), the droplets are moving towards the center of the channel. 
The shear flow stretches and deforms the droplets and surfactant shifts towards the droplets tips, Fig.~\ref{tpsi}(a)-(b). 
Later on, $t=2.0$, the droplets get closer; a thin liquid film separates the two droplets. 
The high strain rate regions, Fig.~\ref{tphi}(c)-(d), highlight the liquid film draining process: as the droplets get closer, the carrier fluid is squeezed out. 
Surfactant further accumulates at the droplets tips, increasing droplet deformation, Fig.~\ref{tpsi}(c)-(d). 
Up to this stage there is no appreciable difference between the two cases. 
At $t=3.0$, Fig.~\ref{tphi}(e)-(f), surfactant effects can be appreciated: for case C the liquid film is thicker and the draining rate is lower (lower $S_x$ magnitude) with respect to case A. 
In Fig.~\ref{tphi}(e), a region with strong negative strain can be observed in the middle of the gap: this region is characterized by a lower pressure which draws the interface towards coalescence.
For case A the liquid film drains before $t=4.0$ and the droplets coalesce, Fig.~\ref{tphi}(g); after the coalescence, surface tension reshapes the new droplet and surfactant is redistributed over the interface, Fig.~\ref{tpsi}(g). 
Conversely, surfactant  prevents coalescence in case C, increasing the droplet deformability and generating tangential stresses at the interface; these two combined effects hinder the liquid film draining.
\subsubsection{Deformation during the droplet-droplet interaction}
\begin{figure}
\center
\setlength{\unitlength}{0.0025\columnwidth}
\begin{picture}(400,460)
\put(30,0){\includegraphics[width=0.80\columnwidth, keepaspectratio]{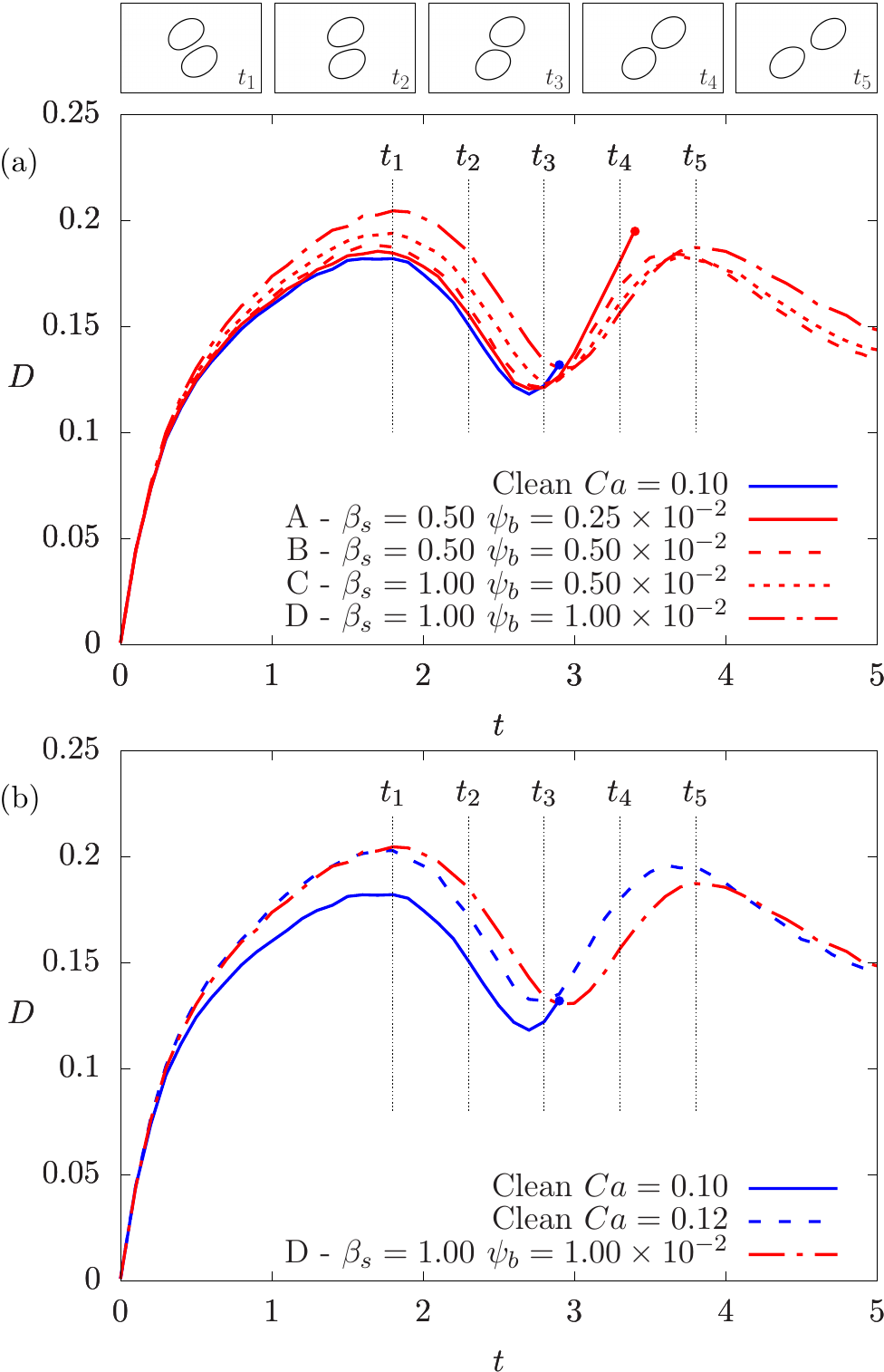}}
\end{picture}
\caption{Deformation parameter $D$ of the droplets during the interaction. 
Panel (a) compares the results obtained at $Ca=0.10$ for different $\psi_b$ and $\beta_s$. 
As $\beta_s$ and $\psi_b$ increase, higher values of the deformation parameter $D$ can be observed.
At $t=2.9$, the two clean droplets coalesce (blue dot); likewise, for case A (red dot) at $t=3.5$. 
For cases C-D-E, droplets do not coalesce and the generic positions of the droplets are reported in the mini-panels on top of the figure.
In Panel (b), the case D (highest amount of surfactant) is compared against a simulation of two clean droplets with an equivalent deformability of case D, thus with an higher capillary number, $Ca=0.12$ (same effective capillary, $Ca_e$ of the surfactant-laden case).}
\label{ddd}
\end{figure}
To investigate the role played by the surfactant in the deformation, we compute the deformation parameter $D$ of the two droplets.
The time evolution of $D$ during the interaction for the cases A-B-C-D and Clean (see Fig.~\ref{map}) have been compared in Fig.~\ref{ddd}(a).
Since the two droplets evolve over time in the same way, only the deformation parameter of one of the droplet has been plotted.
On the top of the figure, five mini-panels show the generic configuration of the system at different times.
The droplets, initially circular ($D=0$), start to deform and move according to the shear flow. 
After $t=1.5$, the droplets are close enough and start to interact; the deformation parameter reaches a maximum, $t=t_1$.
Then, the presence of the neighbouring droplet leads to a reduction of the deformation parameter $D$ that reaches a minimum for $t \simeq t_3$.
The minimum of $D$ is smaller for the Clean case (blue solid line) and increases when $\psi_b$ and/or $\El$ are increased (from case A to case D).
This stage of the interaction is crucial in determining its outcome: indeed, higher deformations slow down the draining of the thin liquid film.
After $t\simeq t_3$, for the Clean and the A cases, the droplets coalesce (blue/red dots) and $D$ is not anymore computed.
By opposite, for cases B-C-D, after $t \simeq t_3$, the two droplets separate; the deformation parameter $D$ increases and reaches a new maximum at $t \simeq t_5$.
After $t \simeq t_5$, $D$ decreases and reaches an asymptotic value ($t > 5$, not reported here).

These results suggest that the surfactant can prevent the coalescence acting on the deformability.
To prove this observation, case D has been compared with an additional case with an equivalent deformability but without surfactant, Fig.~\ref{ddd}(b).
The equivalent $Ca$ has been computed from Eq.~(\ref{sheq}), given the steady-state deformation value for case D, obtaining $Ca=0.12$. 
In Fig.~\ref{ddd}(b), the Clean case at $Ca=0.10$ has been also reported as reference.
Interestingly, we can notice that for this new case, $Ca=0.12$, the droplets do not coalesce.
Comparing the results, we can observe a similar behavior for case D and $Ca=0.12$. 
However, some differences can be noticed: the time evolution of the case D is slightly delayed, this delay is probably due to the tangential stresses at the interface.

Overall, the results of Figs.~\ref{ddd}(a)-(b), confirm the role played by the deformability in determining the interaction outcome.
Surfactant, decreasing the surface tension, increases the deformation of the droplets and hinders the draining of the thin liquid film.
Coalescence is favoured when droplets are less deformed (Clean at $Ca=0.10$ and case A) and is prevented when droplets are more deformed (cases B-C-D and $Ca=0.12$).
The behaviors of the deformation parameter $D$ obtained are in good agreement with previous experiments \cite{guido1998} and numerical studies \cite{bayareh2011,singh2009effects,xu2011numerical}.

\subsubsection{Effect of tangential stresses at the interface}
\label{sec:mar}

\begin{figure}[!h]
\center
\setlength{\unitlength}{0.0025\columnwidth}
\begin{picture}(400,300)
\put(0,0){\includegraphics[width=1.00\columnwidth, keepaspectratio]{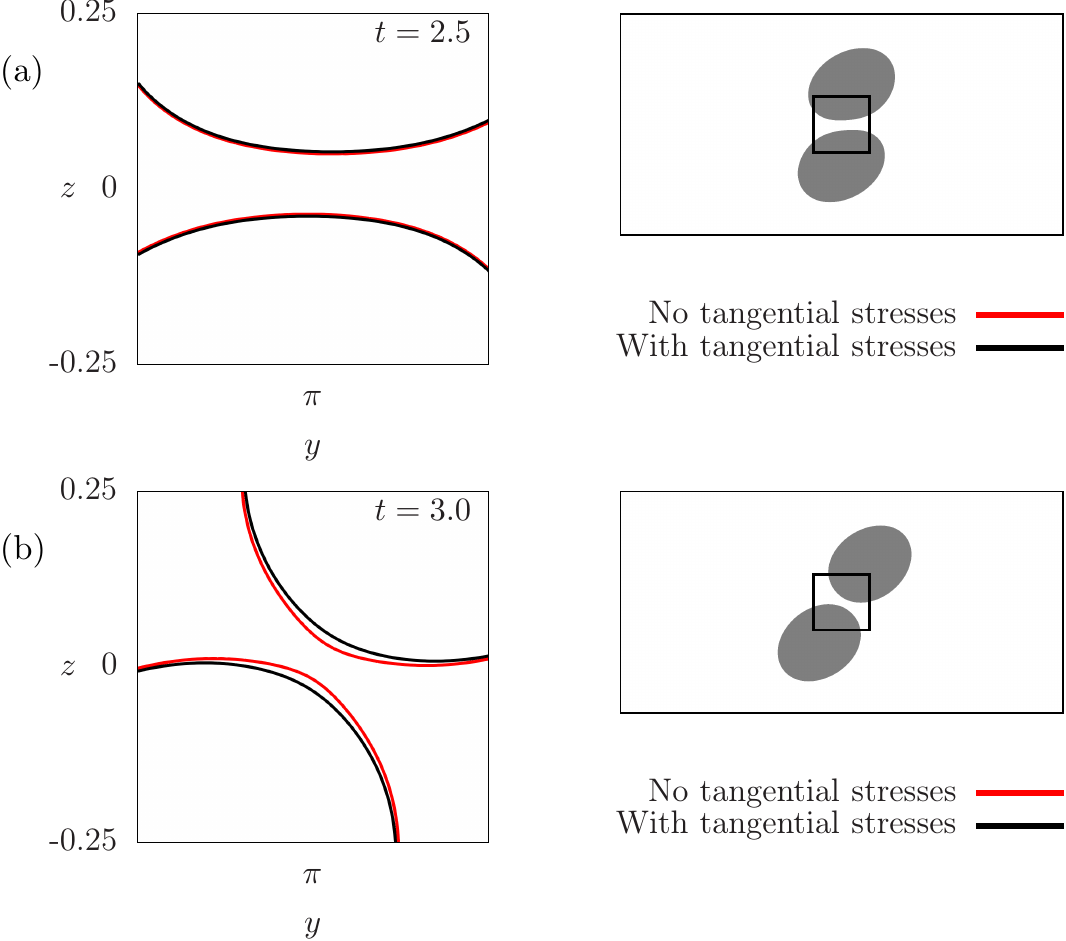}}
\end{picture}
\caption{Instantaneous position of the interface (iso-contour $\phi=0$) at $t=2.5$ in panel (a), and $t=3.0$ in panel (b) for the simulation B. 
The panels show a close-up view of the area highlighted by the black rectangle in the right column.
The two cases are represented in black (tangential stresses considered) and red (tangential stresses neglected).
When neglecting the tangential stresses the draining is faster and the two droplets are closer; this favours coalescence.
By opposite, when tangential stresses are considered, the draining is slower and coalescence is hindered.}
\label{int}
\end{figure}

To understand the effect of tangential stresses (generated by surface tension gradients) on the outcome of the interaction, case B has been recomputed considering only the non-uniform capillary stresses and neglecting the tangential ones.
In particular, the second term on the right hand side of Eq.~(\ref{tanstress}) has been neglected.
This term identifies the tangential stresses generated by surface tension gradients (and thus by surfactant concentrations gradients) at the interface.
When considering both  capillary and tangential contributions of Eq.~(\ref{tanstress}), the simulation (corresponding to case B) leads to a non-coalescence, Fig.~\ref{map}. A complete different outcome (coalescence) is obtained when recomputing the same case but neglecting tangential stresses.
To appreciate the different dynamics of the interaction, the instantaneous positions of the interface ($\phi=0$) when the stresses are considered (black) and neglected (red) have been compared in Fig.~\ref{int}(a)-(b). 
At $t=2.5$, Fig.~\ref{int}(a), the difference between the position of the interfaces is small: the droplets are slightly closer when tangential stresses are neglected.
Later on, $t=3.0$, Fig.~\ref{int}(b), the difference between the two interfaces is much larger.
Neglecting tangential stresses allows the droplets to get closer, thanks to the higher liquid film draining rate. 
The draining rate determines whether the droplets will coalesce; a lower draining rate (hindered by tangential stresses) prevents coalescence.

\begin{figure}
\center
\setlength{\unitlength}{0.0025\columnwidth}
\begin{picture}(400,150)
\put(0,0){\includegraphics[width=1.00\columnwidth, keepaspectratio]{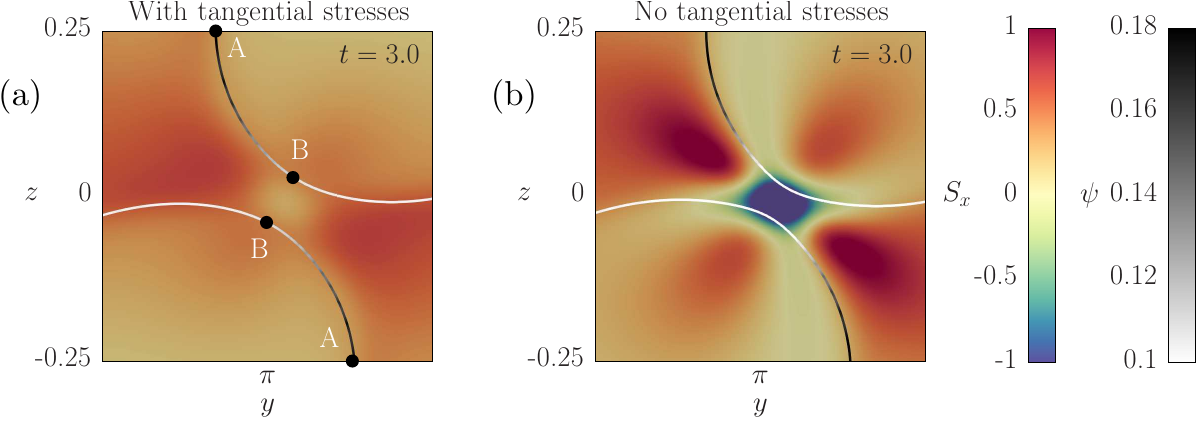}}
\end{picture}
\caption{Contour map of the strain rate $S_x$ for the simulation B.
In panel (a) the simulation is performed considering the tangential stresses while in panel (b) these stresses are neglected.
The interface, iso-contour $\phi=0$, is colored by the surfactant concentration (white-low and black-high). 
The tangential stresses hamper the draining of the thin liquid film and suppress the formation of high strain rate magnitude regions.
When neglected, panel (b), the draining process is faster and regions with high magnitude of $S_x$ are present (dark-red and blue areas).}
\label{streamar}
\end{figure}

To shed some light on this mechanism, the instantaneous strain rate $S_x$ and the surfactant concentration at the interface are shown in Fig.~\ref{streamar}.
Tangential stresses drive fluid along the interface from a region with high surfactant concentration (point A) to a region with lower surfactant concentration inside the gap (point B), Fig.~\ref{streamar}(a).  This flow opposes to the liquid film draining, hampering it. 
The importance of tangential stresses can be appreciated comparing the two panels of Fig.~\ref{streamar}, as they prevent the formation of high strain rate magnitude regions: in the liquid film the strain rate has a value closer to the mean strain rate (due to the shear flow), $S_x=0.5$ (Fig.~\ref{streamar}(a)). 
By opposite, high strain rate magnitude regions appear in the gap when Marangoni stresses are neglected, Fig.~\ref{streamar}(b), indicating a stronger gap draining phenomenon.

Summarizing, surfactant helps in preventing coalescence with two mechanisms: $(i)$ an increase of the droplet deformability (lower surface tension) increases the thickness of the thin liquid film $(ii)$ tangential stresses at the interface drive fluid inside the liquid film. 
Both mechanisms act in the same direction, hindering the liquid film draining and, thus, droplets coalescence.
The results obtained are in agreement with experimental \cite{dai2008,guido1998,leal2004flow,pan2016,wang2016} and numerical \cite{xu2011numerical} results for head-on and offset collision of droplets.

\subsection{Swarm of surfactant-laden droplets in a turbulent channel flow}
\label{swarm}

\begin{figure}[t]
\centering
\setlength{\unitlength}{0.0025\columnwidth}
\begin{picture}(400,150)
\put(52,11){\scriptsize{$x$}}
\put(22,11){\scriptsize{$y$}}
\put(42,26){\scriptsize{$z$}}
\put(0,120){(a)}
\put(200,120){(b)}
\put(0,0){\includegraphics[width=1.00\columnwidth, keepaspectratio]{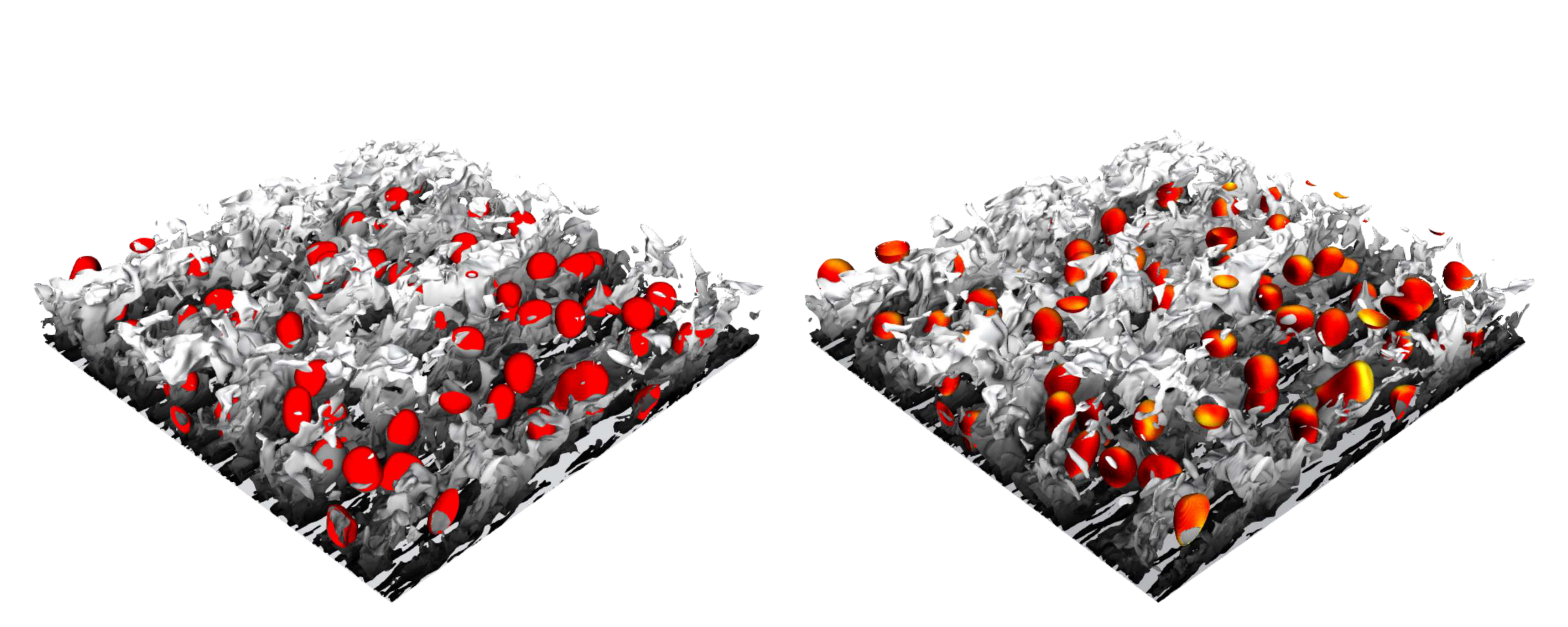}}
\end{picture}
\caption{Swarm of clean, panel (a), and surfactant-laden droplets, panel (b), released in a turbulent channel flow ($Re_\tau=590$).
The droplets interfaces are identified by the iso-contour $\phi=0$. For the surfactant-laden case, panel (b), the interface is colored by the surfactant concentration $\psi$ (red-low and yellow-high).
The turbulent structures are highlighted using the iso-surface of the streamwise velocity fluctuations $u'=1.0$ and colored by the distance from the wall (dark: bottom wall, white: center).
For visualization purposes, only part of the domain is reported. In particular, only the bottom part (from $z=-1$ up to $z=0$) and half of the streamwise length (from $x=0$ to $x=2\pi$) is showed. Both the snapshots refer to $t=0.5$.}
\label{turb}
\end{figure}
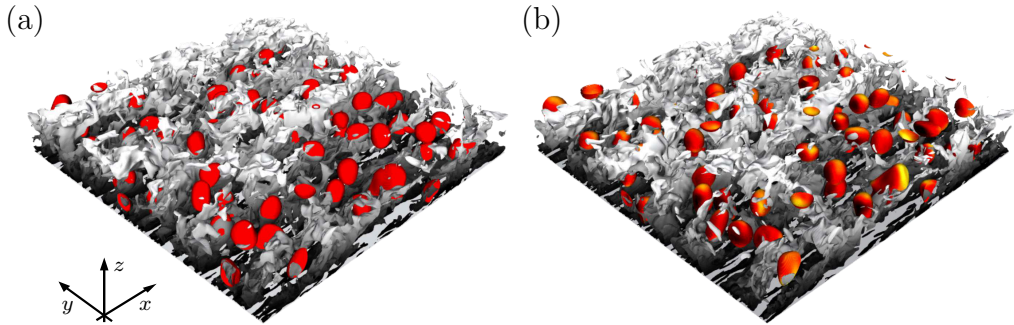

The capabilities of the computational model to handle turbulent flows and large grids have been tested by simulating the dynamics of 256 droplets in a turbulent channel flow; a clean and a surfactant-laden case have been considered.
The computational domain has dimensions $L_x \times L_y \times L_z=4 \pi \times 2 \pi \times 2$ and has been discretized using $N_x \times N_y \times N_z=1024 \times 512 \times 513$ collocation points along the streamwise, spanwise and wall-normal directions. 
The spherical droplets, diameter $d=0.4$, are released in a turbulent flow field at a shear Reynolds number $Re_\tau=590$. 
The initial flow field was obtained from a direct numerical simulation of a single phase flow at $Re_\tau=590$ in a closed channel geometry. 
Once a statistical steady-state is reached, the flow field is saved and used as an initial condition for the droplet-laden turbulent flow.
An array of 256 spherical droplets is initialized in the channel; the phase field equilibrium profile, Eq.~(\ref{eqprofphi}), is imposed at the interface of the droplets.
For the surfactant-laden case, the surfactant concentration field $\psi$ is also initialized with its equilibrium profile, Eq.~(\ref{eqprofpsi}); a surfactant bulk concentration $\psi_b=1.00 \times 10^{-2}$ and an elasticity number $\El=1.00$ have been used.
For both the cases, the Weber number (based on the surface tension of a clean interface, $\sigma_0$) is set to $We=1.50$.

After the release, each of the droplets interacts with the surrounding flow and with the other droplets.
Two qualitative views of the system, one for the clean and one for the surfactant-laden case have been reported in Fig.~\ref{turb}(a)-(b).
Panel (a) refers to the clean case (absence of surfactant), whereas panel (b) to the surfactant-laden one.
In both panels, only a part of the domain is reported, specifically the bottom part of the channel (from $z=-1$ to $z=0$) and half of the streamwise length (from $x=0$ to $x=2\pi$).
The droplets are identified by the iso-contour $\phi=0$ and colored in red (clean) and by the local surfactant concentration (surfactant-laden, red-low and yellow-high).
To highlight the droplet-turbulence interactions, the iso-contour of the streamwise fluctuations $u'=1.0$ is also reported.
The structures are colored by their distance from the bottom wall (dark: bottom wall, white: center).
From these qualitative pictures we can appreciate the complexity of the flow; in addition it can be noticed how the surfactant distribution, panel (b), is not only influenced by the local curvature but also droplet motion and turbulence affect its distribution at the interface.

\begin{figure}[t]
\centering
\setlength{\unitlength}{0.0025\columnwidth}
\begin{picture}(400,150)
\put(40,0){\includegraphics[width=0.75\columnwidth, keepaspectratio]{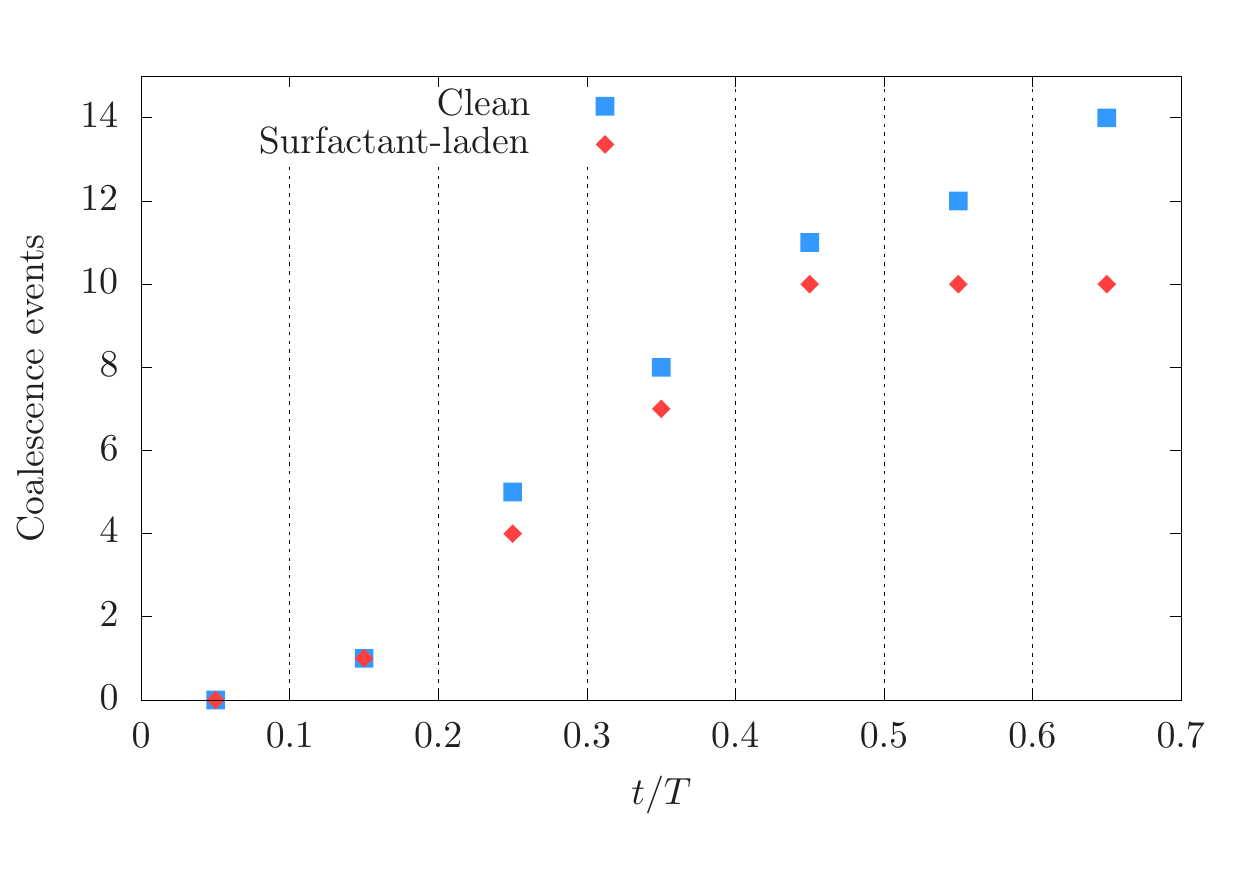}}
\end{picture}
\caption{Number of coalescence events occurring in a small time window (10\% of the channel turnover time, $T$) over time for the droplet-laden turbulent flow.
The blue squares refer to the clean case (absence of surfactant) whereas the red diamonds refer to the surfactant-laden case.
Surfactant seems to be effective even when a turbulent flow is considered, reducing the number of coalescence events with respect to the clean case.}
\label{crate}
\end{figure}

To give a more quantitative indication of the surfactant effect when a turbulent flow is considered, the number of coalescence events occurred in these first stages of the simulation is reported in Fig.~\ref{crate}.
Specifically, we report the number of coalescence events occurred in a small time window $\Delta t=0.1$ over time, in blue for the clean case and red for the surfactant-laden one.
The results show a reduction of the number of coalescence events when surfactant is taken into account.
These observations are in agreement with previous findings; surfactant, increasing the deformability and introducing tangential stresses, can prevent the coalescence even in a turbulent flow field.

Lastly, we would like to remark that these results are reported only to highlight the capabilities of the method; further analysis are out of the scope of this work.

\section{Conclusions}
\label{sec: conclusion}

In this work a modified phase field method for the simulation of surfactant-laden turbulent flows has been presented. 
The modifications introduced, together with the numerical scheme adopted, improve the well-posedness and the flexibility of the method, making it well-suitable for the simulation of surfactant-laden turbulent flows.
In particular, compared to previous works, the unphysical behavior of the interfacial layer has been circumvented by removing the surfactant contribution in the phase field chemical potential. This contribution could change the thickness of the interfacial layer; removing it restores the correct behavior of the interfacial layer and relaxes the grid requirements \cite{Yun2014}.
The surface tension forces in the Navier-Stokes equation are now computed using a geometrical approach (instead of the commonly adopted thermodynamic one \cite{Engblom2013,Laradji1992,toth2015}), together with an equation of state that accounts for the surfactant effect.
This modification improves the flexibility of the method (arbitrary choice of the equation of state) and allows us to distinguish among the different effects introduced by surfactant (average surface tension decrease, non-uniform capillary forces and tangential stresses at the interface \cite{StoneL_1990}).
In addition, the pseudo-spectral discretization and the parallelization scheme adopted lead to an approach that can handle large-scale simulations (billion of grid points) and runs efficiently on a large number of processes (tested up to 65k tasks).

The proposed PFM has been extensively tested comparing the results obtained with experimental and analytic data.
Specifically, the method has been first validated considering the deformation of a single droplet in shear flow and then used to study the influence of surfactant on the interaction of two droplets in shear flow.
For the latter case, results show that surfactant, increasing the deformability and introducing tangential stresses at the interface, can hinder coalescence.
Lastly, the capabilities of the method in handling large-scale simulation and complex phenomena have been tested considering a 3D simulation of a swarm of surfactant-laden droplets in turbulence.
Preliminary results show that, even when a turbulent flow is considered, surfactant is able to prevent coalescence (reduced number of coalescence events).

Overall, the proposed modified phase field method aims to overcome most of the limitations of the current approaches and is well-suitable for large-scale simulations of turbulent multiphase flows.
From a computational point of view, the approach has an optimal scalability and, in terms of computational cost (assumed to be approximately proportional to the number of fast Fourier transforms per time step), the simulation of a surfactant-laden flow is about four times more demanding than that of a single phase flow on the same grid.

\section*{Acknowledgments}
CINECA supercomputing center (Bologna, Italy) and Vienna Scientific Cluster (Vienna, Austria) are gratefully acknowledged for generous allowance of computer resources.
AR and GS thankfully acknowledge support from FSE EUSAIR-EUSALP, and from the Erasmus Traineeship program, both coordinated by University of Udine

\vspace{1cm}

Declarations of interest: none
\vspace{0.5cm}

This research did not receive any specific grant from funding agencies in the public, commercial, or not-for-profit sectors.

\section*{References}
\bibliographystyle{plainnat}
\bibliography{totalbib.bib}

\appendix

\section{Non-dimensionalization}
\label{sec: nondim}
In this section the non-dimensionalizing procedure will be presented, together with the dimensionless numbers introduced. Dimensional variables will be denoted here as $\widetilde{\theta}$, while the dimensionless ones as $\theta$, being $\theta$ a generic variable.
The dimensional free energy functional is:
\begin{equation}
\widetilde{\mathcal{F}}=\int_\Omega\left(\widetilde{f}_0+\widetilde{f}_{mix}+\widetilde{f}_\psi+\widetilde{f}_1+\widetilde{f}_\Ex\right)d\Omega
\end{equation}
The dimensional phase field variable is defined as: $\widetilde{\phi}=\sqrt{\beta/\alpha}\phi$. The dimensional phase field free energy is ($\alpha$, $\beta$ and $\kappa$ are the parameters of the Cahn-Hilliard model \cite{Jacqmin1999}):
\begin{equation}
\begin{aligned}
\widetilde{f}_0+\widetilde{f}_{mix}&=\frac{\alpha}{4}\left(\widetilde{\phi}-\sqrt{\frac{\beta}{\alpha}} \right)^2 \left(\widetilde{\phi}-\sqrt{\frac{\beta}{\alpha}} \right)^2 +\frac{k}{2}|\nabla\widetilde{\phi}|^2=\\
&=\frac{\beta^2}{\alpha}\left[\frac{1}{4}\left(\phi-1 \right)^2 \left(\phi+1 \right)^2 +\frac{\Ch^2}{2}|\nabla\phi|^2\right]
\end{aligned}
\end{equation}
Here we have exploited the relationships $\varepsilon=\sqrt{\kappa/\beta}$ and $\Ch=\varepsilon/h$. $\varepsilon$ is the lengthscale of the interface thickness.

For the entropy decrease term, we have:
\begin{equation}
\begin{aligned}
\widetilde{f}_\psi&=\kappa T \left[\psi\log \psi +(1-\psi)\log(1-\psi)\right]=\\
&=\frac{\beta^2}{\alpha}\PI \left[\psi\log \psi +(1-\psi)\log(1-\psi)\right]
\end{aligned}
\end{equation}
The coefficient $\PI$ is defined as $\PI=\kappa T\alpha/\beta^2$; $T$ is the absolute temperature, while $\alpha$, $\beta$ and $\kappa$ are the same parameters defined for the phase field free energy. The surfactant volume fraction is already a dimensionless quantity from its definition.
The dimensional surfactant adsorption contribution is:
\begin{equation}
\begin{aligned}
\widetilde{f}_1&=-\frac{\kappa}{2}\psi|\nabla\widetilde{\phi}|^2=-\frac{\beta^2}{\alpha}\frac{\Ch^2}{2}\psi|\nabla\phi|^2
\end{aligned}
\end{equation}
Using Model-3 from \cite{Engblom2013} we have instead:
\begin{equation}
\begin{aligned}
\widetilde{f}_1&=-\frac{\alpha}{2}\psi\left(\frac{\beta}{\alpha}-\widetilde{\phi}^2\right)^2=-\frac{\beta^2}{\alpha}\frac{1}{2}\psi(1-\phi^2)^2
\end{aligned}
\end{equation}
Finally, the dimensional surfactant bulk part is:
\begin{equation}
\widetilde{f}_\Ex=\frac{w}{2}\psi\widetilde{\phi}^2=\frac{\beta^2}{\alpha}\frac{1}{2\Ex}\psi\phi^2
\end{equation}
The dimensionless parameter $\Ex$ is defined as $\beta/w$.

The dimensionless free energy functional is defined as: $\mathcal{F}=\widetilde{\mathcal{F}}\alpha/\beta^2$, thus resulting in:
\begin{equation}
\begin{aligned}
\mathcal{F}=&\int_\Omega\left(\frac{1}{4}\left(\phi-1 \right)^2 \left(\phi+1 \right)^2 +\frac{\Ch^2}{2}|\nabla\phi|^2+\right.\\
&\left.+\PI \left[\psi\log \psi +(1-\psi)\log(1-\psi)\right] -\frac{\Ch^2}{2}\psi|\nabla\phi|^2+\frac{1}{2\Ex}\psi\phi^2\right)d\Omega
\end{aligned}
\end{equation}
If we use Model-3 from Engblom \textit{et al.} we get:
\begin{equation}
\begin{aligned}
\mathcal{F}=&\int_\Omega\left(\frac{1}{4}\left(\phi-1 \right)^2 \left(\phi+1 \right)^2 +\frac{\Ch^2}{2}|\nabla\phi|^2+\right.\\
&\left.+\PI \left[\psi\log \psi +(1-\psi)\log(1-\psi)\right] -\frac{1}{2}\psi(1-\phi^2)^2+\frac{1}{2\Ex}\psi\phi^2\right) d\Omega
\end{aligned}
\end{equation}
The dimensional transport equation for the phase field variable is:
\begin{equation}
\frac{\de \widetilde{\phi}}{\de \widetilde{t}}+\mathbf{\widetilde{u}}\cdot\nabla\widetilde{\phi}=\nabla\cdot(\widetilde{\mathcal{M}}_\phi\nabla\widetilde{\mu}_\phi)
\end{equation}
From the non-dimensionalization of the free energy functional we get that the dimensionless chemical potential for the phase variable is defined as: $\mu_\phi=\sqrt{\alpha/\beta^3}\widetilde{\mu}_\phi$. The length scale of the problem is the channel half height $h$ and the velocity scale is the shear velocity $u_\tau$; from these two scales we can define the time scale $h/u_\tau$. The P\'eclet number for the phase variable $\Pe_\phi$ is defined as:
\begin{equation}
\Pe_\phi=\frac{u_\tau h}{\beta\widetilde{\mathcal{M}}_\phi}
\end{equation}
The dimensionless transport equation for the phase variable thus reads:
\begin{equation}
\frac{\de \phi}{\de t}+\mathbf{u}\cdot\nabla\phi=\frac{1}{\Pe_\phi}\nabla^2\mu_\phi
\end{equation}
The dimensional equation for the surfactant volume fraction transport is:
\begin{equation}
\frac{\de \widetilde{\psi}}{\de \widetilde{t}}+\mathbf{\widetilde{u}}\cdot\nabla\widetilde{\psi}=\nabla\cdot(\widetilde{\mathcal{M}}_\psi\nabla\widetilde{\mu}_\psi)
\end{equation}
Here the dimensional mobility $\widetilde{\mathcal{M}}_\psi$ can be rewritten as a reference constant dimensional mobility $\widetilde{m}_\psi$ and a dimensionless variable part $\mathcal{M}_\psi=\psi(1-\psi)$: 
\begin{equation}
\widetilde{\mathcal{M}}_\psi=\widetilde{m}_\psi\psi(1-\psi)=\widetilde{m}_\psi \mathcal{M}_\psi
\end{equation}
We now define the P\'eclet number for the surfactant phase $\Pe_\psi$:
\begin{equation}
\Pe_\psi=\frac{u_\tau h \alpha}{\widetilde{m}_\psi\beta^2}
\end{equation}
The dimensionless transport equation for the surfactant volume fraction thus results in:
\begin{equation}
\frac{\de \psi}{\de t}+\mathbf{u}\cdot\nabla\psi=\frac{1}{\Pe_\psi}\nabla^2\mu_\psi
\end{equation}
Assuming two phases with matched density ($\rho=\rho_1=\rho_2$) and viscosity ($\eta=\eta_1=\eta_2$), the dimensional Navier-Stokes equation is:
\begin{equation}
\rho \left( \frac{\de \widetilde{\mathbf{u}}}{\de\widetilde{t}}+ \widetilde{\mathbf{u}}\cdot\nabla\widetilde{\mathbf{u}} \right)=-\nabla\widetilde{p}+\eta \nabla^2 \widetilde{\mathbf{u}} +\nabla\cdot\left[\widetilde{\overline{\tau}}_c \kappa \sigma_0 f_\sigma(\psi) \right]
\label{eq: nsdim}
\end{equation}
The velocity scale is the shear velocity $u_\tau$ and the length scale the channel half height; the time scale can be obtained by the length and velocity scales. The dimensionless term $f_\sigma(\psi)$ accounts for the surface tension reduction due to the presence of surfactant; it corresponds to the dimensionless Langmuir EOS for the surface tension. The dimensional pressure is defined as $\widetilde{p}=\rho u_\tau^2 p$; the Korteweg stress tensor is made dimensionless by:
\[
\widetilde{\overline{\tau}}_c= \frac{\beta}{\alpha h^2}\overline{\tau}_c
\]
Two dimensionless groups can be identified: the shear Reynolds number $\Re_\tau$ and the Weber number $\We$.
\begin{equation}
\Re_\tau=\frac{\rho u_\tau h}{\eta}
\end{equation}
\begin{equation}
\We=\frac{\rho u_\tau^2 h}{\sigma_0}
\end{equation}
The surface tension is defined as the integral of the specific energy stored in the interface \cite{YUE2004} and it results in:
\begin{equation}
\widetilde{\sigma}=\sigma_0 f_\sigma(\psi)=\frac{\sqrt{8}}{3}\frac{k^{\frac{1}{2}}\beta^\frac{3}{2}}{\alpha}f_\sigma(\psi)
\label{eq: sigmadim}
\end{equation}
The dimensionless surface force term is:
\[
\frac{h}{\rho u_\tau^2} \frac{\kappa}{h} \frac{\beta}{\alpha h^2}\nabla\cdot(\overline{\tau}_c f_\sigma(\psi))
\]
where the first part comes from the non-dimensionalization of the left hand side of the Navier-Stokes equations. From its coefficient we have:
\[
\frac{h}{\rho u_\tau^2} \frac{\kappa}{h} \frac{\beta}{\alpha h^2}=\frac{\sqrt{\frac{\kappa}{\beta}}}{h}\frac{\sqrt{\beta^3\kappa}}{\alpha}\frac{1}{h\rho u_\tau^2}=\Ch \frac{3}{\sqrt{8}}\sigma_0 \frac{1}{h\rho u_\tau^2}=\frac{3}{\sqrt{8}}\frac{\Ch}{\We}
\]
We can now write the dimensionless Navier-Stokes equations:
\begin{equation}
\frac{\de\mathbf{u}}{\de t}+\mathbf{u}\cdot\nabla\mathbf{u}=-\nabla p +\frac{1}{\Re_\tau}\nabla^2\mathbf{u}+\frac{3}{\sqrt{8}}\frac{\Ch}{\We}\nabla\cdot(\overline{\tau}_c f_\sigma(\psi))
\end{equation}



%
%
%

\end{document}